\documentclass[acmsmall, nonacm, authorversion]{acmart}
\usepackage{hyperref}
\usepackage{amsmath,amsfonts}
\usepackage{algorithmic}
\usepackage{graphicx}
\usepackage{textcomp}
\usepackage{xcolor}
\usepackage{url}
\usepackage{array}
\usepackage{subcaption}
\usepackage{listings}
\usepackage{framed}
\usepackage{tabularx}
\usepackage{caption}
\usepackage[utf8]{inputenc}

\def\BibTeX{{\rm B\kern-.05em{\sc i\kern-.025em b}\kern-.08em
    T\kern-.1667em\lower.7ex\hbox{E}\kern-.125emX}}

\lstdefinestyle{clean}{
  basicstyle=\footnotesize\ttfamily,
  frame=none,
  numbers=none,
  columns=fullflexible,
  keepspaces=true,
  showstringspaces=false,
  breaklines=true,
  keywordstyle=\bfseries\color{black}, %
  commentstyle=\itshape,               %
  stringstyle=\itshape                 %
}

\lstdefinestyle{compactcode}{
  basicstyle=\footnotesize\ttfamily,
  columns=fullflexible,
  keepspaces=true,
  showstringspaces=false,
  numbers=left,
  numberstyle=\scriptsize,
  numbersep=6pt,
  aboveskip=2pt,
  belowskip=2pt,
  frame=single,
  framesep=3pt,
  xleftmargin=12pt
}

\definecolor{keywordcolor}{rgb}{0.8, 0.0, 0.0}  %
\definecolor{loopcolor}{HTML}{1F77B4}    %
\definecolor{updatecolor}{rgb}{0.5, 0.0, 0.5}  %

\lstdefinelanguage{CustomC}{
    keywords={void, int, for, bool}, %
    keywordstyle=\color{keywordcolor}\bfseries,
    morekeywords=[2]{i, j, k}, %
    keywordstyle=[2]\color{loopcolor}\bfseries,
    morekeywords=[3]{update, temp, C}, %
    keywordstyle=[3]\color{updatecolor}\bfseries,
    sensitive=true,
    basicstyle=\ttfamily\small,
    morecomment=[l]{//},
    commentstyle=\color{gray}\itshape,
    morestring=[b]",
}

\definecolor{ircomment}{RGB}{0,128,0}
\definecolor{irkeyword}{RGB}{0,0,180}
\definecolor{irtype}{RGB}{128,0,128}

\lstdefinelanguage{LLVMIR}{
  morekeywords={
    define, declare, target, datalayout, triple,
    void, i1, i8, i16, i32, i64, float, double,
    ret, br, call, phi, add, sub, mul, sdiv, udiv, 
    load, store, alloca, getelementptr, icmp, fcmp, 
    eq, ne, sgt, sge, slt, sle, ugt, uge, ult, ule,
    and, or, xor, shl, lshr, ashr, sext, zext, fadd, 
    fsub, fmul, fdiv, fptrunc, fpext, sitofp, fptosi, 
    bitcast, ptrtoint, inttoptr, select, unreachable
  },
  sensitive=true,
  morecomment=[l];,
  morestring=[b]",
  commentstyle=\color{ircomment}\ttfamily,
  keywordstyle=\color{irkeyword}\bfseries,
  stringstyle=\color{irtype},
  basicstyle=\ttfamily\footnotesize,
  showstringspaces=false,
  columns=fullflexible,
  escapeinside={(*@}{@*)}
}

\lstdefinelanguage{CFG}{
  morekeywords={::=, |, if, from, to, in, for, while},
  sensitive=false,
  morecomment=[l]{//},
  morestring=[b]"
}

\newcommand{\myinfrule}[3]{%
  \begin{tabular}[c]{@{}c@{}}
    $#1$ \\[-1pt]
    \hline
    $#2$
  \end{tabular}%
  \,\textnormal{ (#3)}%
}

\newcommand{\rulesep}{%
  \hspace{1.5em}\allowbreak
  \penalty0
  \hspace{0pt}
}

\setcopyright{cc}
\setcctype{by}
\acmDOI{10.1145/3839483}
\acmYear{2026}
\acmJournal{PACMPL}
\acmVolume{10}
\acmNumber{OOPSLA2}
\acmArticle{351}
\acmMonth{10}
\acmSubmissionID{oopslab26main-p705-p}
\received{2026-03-23}
\received[accepted]{2026-08-06}

\begin{document}
\title{Symbolic Basic Block Profiling for Machine Learning Kernels}



\author{Jingyu Qiu}
\orcid{0009-0004-4742-1248}
\affiliation{%
  \institution{University of Rochester}
  \city{Rochester}
  \country{USA}
}
\email{jqiu9@ur.rochester.edu}

\author{Rongcui Dong}
\orcid{0000-0003-0146-8417}
\affiliation{%
  \institution{University of Rochester}
  \city{Rochester}
  \country{USA}
}
\email{rdong3@cs.rochester.edu}

\author{Sreepathi Pai}
\orcid{0000-0002-3691-7238}
\affiliation{%
  \institution{University of Rochester}
  \city{Rochester}
  \country{USA}
}
\email{sree@cs.rochester.edu}

\begin{abstract}
Current basic block profiling techniques obtain the count of executions of each basic block in a program using dynamic instrumentation. These profiling counters create  runtime overheads and also require the execution of the program, which, for large input sizes, can take substantial time. We propose symbolic program profiling that generates symbolic formulae for a basic block's count with inputs as the independent variables. 
Our technique is limited in applicability to a certain class of programs, namely machine learning (ML) kernels.
We implement our technique in the LLVM compiler and evaluate it on 78 ML operators from 50 different ML models.
These operators are generated by TVM, a machine learning compiler.
Our symbolic profiles deliver exactly the same results as dynamic instrumentation for 73 out of 78 kernels with a median speedup of 15093$\times$.
\end{abstract}

\maketitle

\section{Introduction}

Basic block profiling is an important technique in program performance analysis. A profile records the execution counts of each basic block and enables better optimizations by providing dynamic information that is otherwise not available to a compiler. 
Most compilers use these profiles in profile-guided optimization (PGO)~\cite{usingProfileInformation} to improve the quality of the generated code.   
Such optimizations have been adopted in data centers and clouds to reduce operational costs \cite{chenAutoFDOAutomaticFeedbackdirected2016,panchenkoBOLTPracticalBinary2019}. 

Instrumentation-based techniques~\cite{continuousPathAndEdgeProfiling}  insert counters into a program to collect profiles during execution. 
Counters result in exact profiles, but at the cost of runtime overhead incurred by the extra counter operations. 
This overhead can be high and hinders widespread use~\cite{panchenkoBOLTPracticalBinary2019}. 
Sampling-based techniques~\cite{chenTamingHardwareEventSamples, aProgrammableHardwarePathProfiler} use hardware support to reduce runtime overhead at the cost of profile accuracy~\cite{chenTamingHardwareEventSamples}. 
Sampled profiles can miss some opportunities in profile-guided optimization~\cite{chenAutoFDOAutomaticFeedbackdirected2016} and cannot be used for some applications that require exact profiles~\cite{soaresSideChannelElimination, soaresMemorySafeElimination}.
Both techniques also require the execution of the program. 
If the programs take long to run, such as machine learning programs, obtaining these profiles on representative input sizes can dominate the time for collecting these profiles.
Extrapolating from small sizes to larger sizes of inputs is infeasible since machine learning compilers now specialize kernel code generation based on input sizes which are usually fixed in machine learning models. 

Both instrumentation-based and sampling-based profiling techniques work for all programs and have been significantly optimized in prior work~\cite{ballEfficientPathProfiling1996, frenotReducingOverheadExact2024}. 
In our work, we eschew generality and specialize our basic block profiling technique for Machine Learning (ML) workloads. 
Many ML kernels are well-structured with statically-analyzable loop patterns whose executions are fully determined by input matrix/tensor dimensions. 
Thus, the execution frequency of the basic blocks can be calculated solely based on the tensor shapes. 
Our goal is to create a parametric profile for these ML kernels through static analysis. 
This symbolic approach is exact and can be fast -- since only formulae need to be evaluated.
Moreover, it is significantly easier to obtain profiles for different-sized inputs.

Our proposed technique, symbolic basic block profiling, statically generates program profiles that consist of symbolic expressions, which can be evaluated to create a concrete profile when an input is specified. 
In this paper, we introduce a framework for symbolic program profiling. A procedure is first analyzed to extract symbolic expressions that will determine all the basic block counts. These symbolic expressions incorporate, among other things, loop iteration counts and branch taken/non-taken ratios. 
Ultimately, these expressions are rewritten to use only program parameters to determine the basic block count.

We also implement a symbolic profiling tool based on our framework for LLVM IR, which generates a symbolic profile directly from LLVM IR. 
In the experiments, we show that the proposed approach is significantly faster than traditional instrumentation-based techniques in extracting exact program profiles.

This paper is structured as follows. Section~\ref{sec:motivation} introduces symbolic profiles using matrix multiply as an example. Section~\ref{sec:scope} delimits the scope of our technique.
Section~\ref{sec:framework} presents the symbolic profiling framework. 
Section~\ref{sec:impl} describes our implementation in LLVM IR. Section~\ref{sec:evaluation} evaluates the applicability and overhead of the proposed technique on machine learning kernels. We discuss related work in Section~\ref{sec:related}. 
Section~\ref{sec:conclusion} concludes this paper.

\section{Motivation}\label{sec:motivation}

\begin{figure}[tbp]
    \centering

    \begin{subfigure}[b]{0.48\linewidth}
        \centering
        \includegraphics[width=0.9\linewidth]{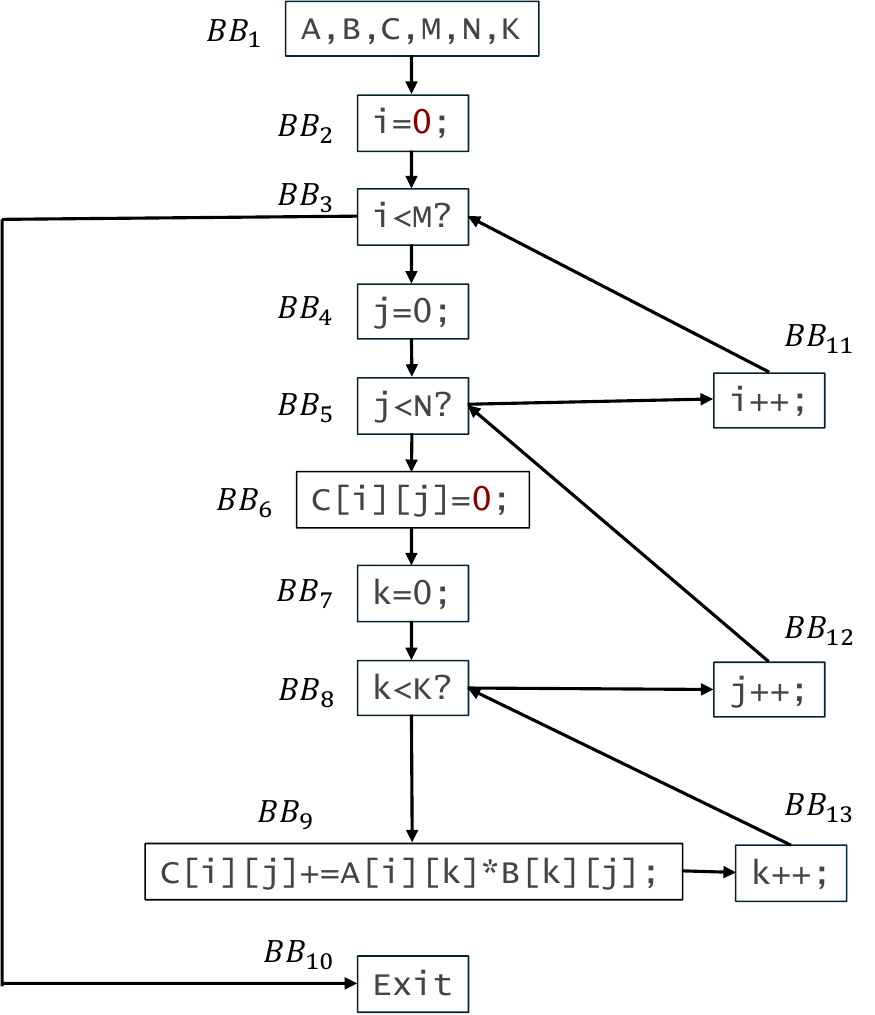}
        \caption{CFG for matrix multiplication kernel in Listing~\ref{lst:naive-matmul}.}
        \label{fig:naive-matmul-cfg}
    \end{subfigure}
    \hfill
    \begin{subfigure}[b]{0.48\linewidth}
        \centering
        \small
        \begin{tabular}{|c|c|}
            \hline
            Block & Expression \\
            \hline
            \( BB_1 \)  & 1 \\
            \( BB_2 \)  & 1 \\
            \( BB_3 \)  & \( M+1 \) \\
            \( BB_4 \)  & \( M \) \\
            \( BB_5 \)  & \( M(N+1) \) \\
            \( BB_6 \)  & \( MN \) \\
            \( BB_7 \)  & \( MN \) \\
            \( BB_8 \)  & \( MN(K+1) \) \\
            \( BB_9 \)  & \( MNK \) \\
            \( BB_{10} \) & 1 \\
            \( BB_{11} \) & \( M \) \\
            \( BB_{12} \) & \( MN \) \\
            \( BB_{13} \) & \( MNK \) \\
            \hline
        \end{tabular}
        \caption{Statically determined expressions for basic blocks.}
        \label{tab:block-expressions}
    \end{subfigure}

    \caption{Control-flow structure and corresponding static expressions for execution counts for the naïve matrix multiplication kernel.}
    \label{fig:matmul-cfg-and-table}
    \Description{The left panel presents a control-flow graph with thirteen basic blocks representing the entry, initialization, and three nested loops of matrix multiplication. The right panel maps each basic block to a symbolic execution-count expression. Representative counts include M plus one for the outer-loop header, M times N plus one for the middle-loop header, M times N times K plus M times N for the inner-loop header, and M times N times K for the multiplication body.}
\end{figure}

We introduce our technique and  demonstrate its advantage over instrumentation-based profiling using an example.
Listing~\ref{lst:naive-matmul} shows the pseudocode for a naive matrix multiplication. %
Line~\ref{line:inputs} defines the function inputs. 
The variables $A$, $B$ and $C$ are 2-D arrays with $A$, $B$ representing the two matrices to be multiplied while the matrix $C$ stores the result. 
The variables $M$, $N$, and $K$ are scalars that define the size of input arrays.
Specifically, the sizes for array $A$, $B$, $C$ are $M$x$K$, $K$x$N$, and $M$x$N$ respectively. Fig.~\ref{fig:naive-matmul-cfg} shows the corresponding control-flow graph (CFG). 
The basic block $BB_1$ represents the function entry which also contains the function inputs. 
Each loop construct decomposes into three basic blocks. 
For example, the outermost loop in line~\ref{line:outter-loop} is made up of three basic blocks $BB_2$, $BB_3$, $BB_{11}$.

\begin{figure}[tbp]
\centering
\begin{lstlisting}[
  style=clean, 
  language=C,
  caption={Naïve matrix multiplication in C-like pseudocode}, label={lst:naive-matmul},
  numbers=left, numberstyle=\tiny, numbersep=6pt,
  xleftmargin=2em,            %
  framexleftmargin=2em,       %
  columns=fullflexible,       %
  breaklines=true,            %
  basicstyle=\ttfamily\footnotesize,
  escapechar=|
]
void matmul_kernel(double **A, double **B, double **C, int M, int N, int K) |\label{line:inputs}|
    for (int i = 0; i < M; ++i) |\label{line:outter-loop}|
        for (int j = 0; j < N; ++j) 
            C[i][j] = 0.0; |\label{line:cij}|
            for (int k = 0; k < K; ++k) 
                C[i][j] += A[i][k] * B[k][j]; |\label{line:cij-end}|
\end{lstlisting}
\Description{C-like pseudocode for matrix multiplication. Three nested loops iterate over dimensions M, N, and K. For each output element, the code initializes C at row i and column j to zero and accumulates the products of corresponding elements from A and B.}
\end{figure}

A simple instrumentation-based basic block profiling for $BB_1$ to $BB_{13}$ would place counters in each block but some of these counters can be eliminated through the \citet{knuthOptimalMeasurementPoints1973} scheme which uses the fact that the sum of the number of transitions into a block equals the sum of transitions out of a block. 
In our example, the conditions inside the branching nodes ($BB_3$, $BB_5$, $BB_8$) are statically unknown because $M$, $N$, $K$ can only be determined at runtime. 
As a result, the compiler may decide to place counters along edges $BB_2\!\rightarrow\! BB_3$, $BB_3\! \rightarrow\! BB_{10}$, $BB_{11}\!\rightarrow\! BB_3$, $BB_{12}\!\rightarrow\! BB_5$, $BB_{13}\! \rightarrow\! BB_8$. 
The remaining counts can be derived from these. 
For example,
the values of $BB_3\! \rightarrow\! BB_4$ can be computed after knowing the values of the other three edges related to $BB_3$.

To derive symbolic counts, we can infer from the source code in Listing~\ref{lst:naive-matmul} that instruction in line~\ref{line:cij} will be executed $M\times N$ times because it is inside a double nested loop. 
Applying this sort of analysis to all the basic blocks, we can get symbolic expressions for all basic blocks in our example as shown in Fig.~\ref{tab:block-expressions}. 
The expression for $BB_6$ reflects our previous reasoning. 
The branching node $BB_3$ gets executed one more time than $BB_4$ due to the loop exit akin to $BB_5$ and $BB_8$.

The symbolic formulae in Fig.~\ref{tab:block-expressions} form a symbolic profile using $M$, $N$, and $K$ as the independent variables. All basic blocks in the naive matmul kernel have a statically determined symbolic expression. 
For simple matmul example, the expressions in Fig.~\ref{tab:block-expressions} are what we would expect from a manual inspection.
However, most production kernels deploy several optimizations -- loop peeling,  unrolling, and vectorization to name a few.
Thus, the kernel may contain basic blocks with non-intuitive symbolic counts. 
For example, when vectorized, the matmul kernel will  have its innermost loop split into two -- a main vectorized loop and an epilogue loop.
The latter handles left-over iterations of the main loop that are not a multiple of the vector size.
For example, from results reported by our tool, we observe an epilogue loop with count
$\texttt{ITE}\!\big(
    (K \bmod 4 = 1),
    \; 0,\;
    M \times N \times ( (K-1) \bmod 4))
  \big)$. This loop count is due to the effect of loop unrolling (of factor 4) along with a loop peeling.

Our symbolic profile offers two advantages. 
First, we can turn it into a concrete profile by replacing the symbolic variables with their runtime values, with constant overhead of simply evaluating the symbolic expressions. Second, compared to an execution profile, the symbolic profile encodes more information, such as dependence based on input size, in the symbolic formulae.

\section{Scope of Symbolic Basic Block Profiling\label{sec:scope}}

The problem of generating symbolic formulae for execution counts in terms of a procedure's input variables has been studied in symbolic complexity analysis (e.g.,~\cite{speed}), empirical complexity analysis (e.g.,~\cite{goldsmithMeasuringEmpiricalComputational2007,plasmaUmassBigO}), and most recently in exact loop bound analysis (ELBA)~\cite{rileyExactLoopBound2025}.

Unlike computational complexity analysis, we seek an exact count. 
Empirical computational complexity analysis~\cite{goldsmithMeasuringEmpiricalComputational2007} \textit{consumes} basic block profiles recorded using instrumentation and uses curve fitting over a pre-determined family of single-variable functions to predict time using a single input variable.
Internally, basic block counts are represented as linear or a polynomial function of a single variable.
This makes it more expensive than purely symbolic techniques and unable to handle counts that involve multiple variables like $M, N, K$ in our example.

Symbolic computational complexity, \`a la SPEED~\cite{speed}, focuses on loops and upper bounds.
Even for the programs we consider, whose loops have easy-to-determine bounds, commonly used control-flow constructs like \texttt{continue} prevent these upper bounds from being tight and hence the counts from being exact.
ELBA, another symbolic technique, discovers linear relationships between values of program variables for single loop programs to discover exact loop bounds.
ELBA can produce loop bounds for complex loops that our technique will be unable to handle, but such loops do not appear in ML programs. 
Translating loop bounds to basic block counts especially in the presence of conditional structures and early exits which do occur in ML programs is not straightforward (Section~\ref{sec:earlyExit}). 
In our analysis, we use LLVM's scalar evolution pass to obtain loop bounds, but ELBA could be used as well.

ML kernels predominantly deal with tensors (multi-dimensional arrays) with arbitrary dimensions (or shapes).
These shape parameters, like $M, N, K$, in our example directly affect the control flow of the program.
Some parameters may not describe the shape of an input tensor, but may also affect control flow, for example, matrix multiply kernels accept a flag denoting a certain input matrix as transposed, changing the order of accessing elements in the loops.

The control flow behavior of some ML kernels also depends on \textit{contents} of tensors, e.g., sorting to find the top-K results. 
Since these contents aren't available during static analysis, 
we deliberately exclude them in this work. 
We denote all procedure parameters other than tensor contents that affect control flow as shape parameters.
Our technique is unable to handle control flow that is dependent on the values of tensors.

To delimit our technique's scope, we introduce Shape-Driven Countable Kernels (SDCK) formally.

\begin{definition}[Execution Count]
Let $P$ be a procedure with parameter set $\Theta$, and let $BB(P)$ denote the set of basic blocks in $P$.  
A procedure's execution is parameterized by an input assignment $\sigma : \Theta \rightarrow \mathcal{V}$, where $\mathcal{V}$ is the domain of program values.

For a basic block $b \in BB(P)$, the \emph{execution count function} $\text{ExecCount}_P(b,\sigma)$ denotes the number of times $b$ executes during the execution of $P$ under input assignment $\sigma$.
\end{definition}

\begin{definition}[Countable on $S$]
Let $P$ be a procedure with parameter set $\Theta$, and let $S \subseteq \Theta$ be a subset of parameters. A basic block $b \in BB(P)$ is said to be \emph{countable on $S$} if there exists a function
$
f_b : \mathcal{V}^{|S|} \rightarrow \mathbb{N}
$
such that for every input assignment $\sigma$,
$
\text{ExecCount}_P(b,\sigma) = f_b(\sigma|_S),
$
where $\sigma|_S$ denotes the restriction of $\sigma$ to the parameters in $S$.
\end{definition} %

\begin{definition}[Shape-Driven Countable Kernel (SDCK)] Additionally, if $S$ is the set of shape parameters, and every basic block of $P$ is countable on $S$, then procedure $P$ is a \emph{Shape-Driven Countable Kernel (SDCK)}.
\end{definition}

Constructing $f_b$ for every basic block in a SDCK procedure is the primary challenge we address in our work.
In practice, $f_b$ can be constructed for a procedure whose loops have shape-determined bounds and whose conditional  statements and early exits have shape-determined frequencies.
Our method constructs $f_b$ based on syntax-induced structure of the procedure, as we detail next.

\section{Framework}\label{sec:framework}

For SDCK procedure, we know each basic block is computable from the shape parameters. 
The naive way to solve this problem would be to construct a function for each basic block but this is not necessary. 
Some basic blocks have the same execution counts, for example in the motivation example Fig.~\ref{fig:naive-matmul-cfg}, $BB_6$ and $BB_7$ will always have the same execution count in all possible executions. 
We develop a region-based counting framework that only relies on region count and branch $trueRatio$ (a number between 0 and 1 that indicates the fraction of times the conditional branch evaluated to true) to derive all basic block counts.

We introduce our representation of programs and related concepts and definitions. Then, we describe how to compute basic block counts.

\subsection{Basic Constructs}\label{sec:language}

\begin{figure*}[tbp]
    \centering
    \begin{subfigure}{0.3\textwidth}
        \centering
        \includegraphics[width=0.25\textwidth]{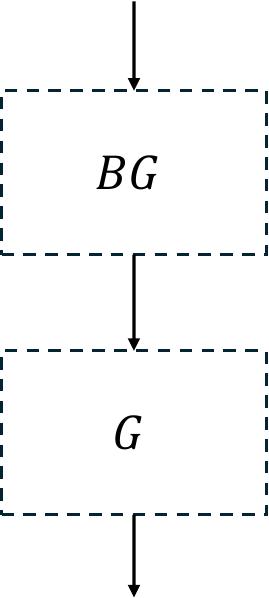}
        \caption{$\langle G \rangle$::= $\langle BG \rangle$ \textbar{} $\langle BG \rangle$ $\langle G \rangle$}
    \end{subfigure}
    \begin{subfigure}{0.3\textwidth}
        \centering
        \includegraphics[width=0.3\textwidth]{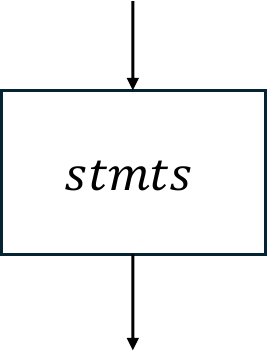}
        \caption{$\langle BB \rangle$::= $stmts$}
    \end{subfigure}
    \begin{subfigure}{0.3\textwidth}
        \centering
        \includegraphics[width=0.6\textwidth]{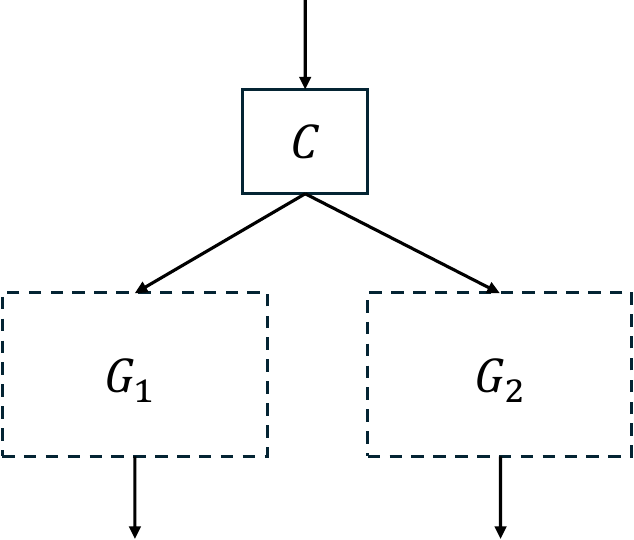}
        \caption{$\langle BR \rangle$::= if ($C$) $\langle G_1 \rangle$, $\langle G_2 \rangle$}
    \end{subfigure}
    \begin{subfigure}{0.4\textwidth}
        \centering
        \includegraphics[width=0.55\textwidth]{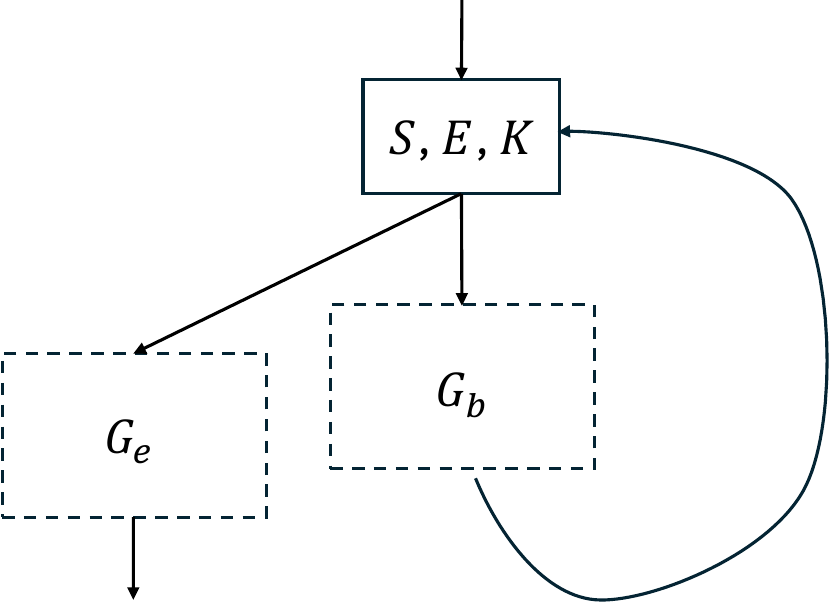}
        \caption{$\langle AL \rangle$::= from $S$ to $E$ in $K$ for $\langle G_b \rangle$ $\langle G_e \rangle$}
    \end{subfigure}
    \begin{subfigure}{0.4\textwidth}
        \centering
        \includegraphics[width=0.55\textwidth]{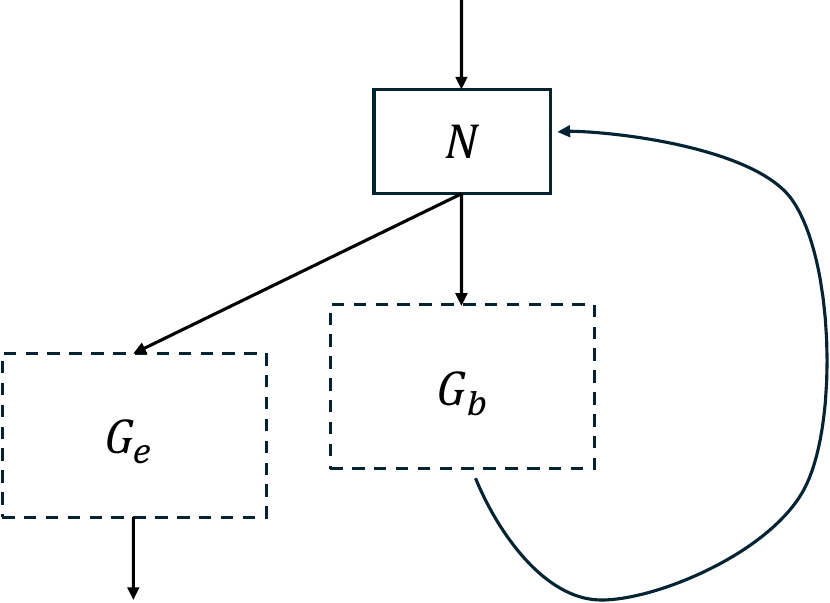}
        \caption{$\langle GL \rangle$::= while ($N$) $\langle G_b \rangle$, $\langle G_e \rangle$}
    \end{subfigure}
    \caption{Visualization of the graph language. Edges represent the control flow. Nodes with solid lines are the basic building blocks which cannot be further expanded. Nodes with dashed lines represent subgraphs that can be further expanded to more basic nodes. (a) shows the production rule of graph $\langle G \rangle$, which consists of a sequence of one or more basic graphs $\langle BG \rangle$. (b) shows the structure of a basic block $\langle BB \rangle$, which only contains some statements. (c) is branch $\langle BR \rangle$, it contains one branching node with the condition $C$, followed by two outgoing subgraphs $\langle G_1 \rangle$ and $\langle G_2 \rangle$, representing two different flows depending on the condition $C$. (d) represents affine loop $\langle AL \rangle$, which consists of a loop entry, a loop body subgraph $\langle G_b \rangle$, and a loop exit subgraph $\langle G_e \rangle$, the trip count of loop body is determined by the $S$, $E$, $K$ tuple. (e) is similar to (d) except for the loop condition $N$.}
    \label{fig:graph-visual}
    \Description{Five control-flow diagrams illustrate the graph-language constructs. A graph is a vertical sequence of basic graphs. A basic block is a solid statement node. A branch condition splits control into two expandable subgraphs. An affine loop contains body and exit subgraphs and a back edge controlled by its start, end, and step values. A general loop has the same structure but is controlled by a general loop condition. Solid boxes denote indivisible nodes, while dashed boxes denote expandable subgraphs.}
\end{figure*}

Rather than defining our analysis directly on the CFG, we use a program representation that lies between high-level programming language constructs and the low-level CFG.  
This simplifies our presentation and allows us to exploit the high-level semantics like loop structures, which are obscured at the CFG level, though which can be recovered using standard compiler techniques~\cite{polygeist}. At the same time, we retain enough of the low-level CFG to make it easier to relate the representation to execution frequency of each basic block. Fig.~\ref{fig:cfg-grammar} contains the grammar for our representation while Fig.~\ref{fig:graph-visual} visualizes each construct as a graph component. 
A program is represented as a graph $\langle G \rangle$. A graph consists of one or more smaller basic graphs $\langle BG \rangle$. 
There are four constructs of basic graphs: basic block $\langle BB \rangle$, branch $\langle BR \rangle$, affine loop $\langle AL \rangle$, and general loop $\langle GL \rangle$. 

\begin{figure}[tbp]
\centering
\small
\begin{minipage}{0.95\linewidth}
\[
\begin{array}{rcl}
\langle G \rangle & ::= & \langle BG \rangle \mid \langle BG \rangle\ \langle G \rangle \\
\langle BG \rangle & ::= & \langle BB \rangle \mid \langle BR \rangle \mid \langle AL \rangle \mid \langle GL \rangle \\
\langle BB \rangle & ::= & \text{statements} \\
\langle BR \rangle & ::= & \text{if } (C)\ \langle G_1 \rangle,\ \langle G_2 \rangle \\
\langle AL \rangle & ::= & \text{from } S \text{ to } E \text{ in } K \text{ for } \langle G_b \rangle,\ \langle G_e \rangle \\
\langle GL \rangle & ::= & \text{while } (N) \ \langle G_b \rangle,\ \langle G_e \rangle
\end{array}
\]
\end{minipage}
\caption{Context-Free Grammar for $\langle G \rangle$}
\label{fig:cfg-grammar}
\Description{A context-free grammar defining the graph representation. A graph is either one basic graph or a basic graph followed by another graph. A basic graph is a basic block, branch, affine loop, or general loop. The productions also define statements within basic blocks, two successor graphs for branches, and body and exit graphs for both loop types.}
\end{figure}

Basic block $\langle BB \rangle$ roughly corresponds to basic blocks in standard CFGs and contains a sequence of statements that are not branches $\langle BR \rangle$ or loops ($\langle AL \rangle$, $\langle GL \rangle$).
However, our basic blocks $\langle BB \rangle$ may contain high-level control statements including subroutine calls and early exit statements such as \texttt{continue}, \texttt{break}, and \texttt{return} with their usual semantics. We provide more details in Section~\ref{sec:earlyExit}.

Branch $\langle BR \rangle$ represents high-level if-statement. It contains one condition and two destination nodes one of which will be taken depending on the value of the condition. 
We use two high-level loop constructs in our representation. 
Affine loops $\langle AL \rangle$ represents loops whose number of iterations through normal exit can be determined by the three variables in the construct, where $S$ stands for start; $E$ stands for end; and $K$ is iteration step. 
Affine loops are very common in ML computation kernels. 
In addition to affine loops, general loops $\langle GL\rangle$ are used to represent other forms of loops to keep our representation general.
Unlike affine loops, $\langle GL\rangle$ has more complex loop patterns, whose iteration count may require specialized analyses to obtain.
We use $N$ to represent the arbitrary loop guards that control the loop iteration.
For both affine loop $\langle AL \rangle$ and general loop $\langle GL \rangle$, $\langle G_b \rangle$ represents loop body and $\langle G_e \rangle$ represents the part of the program after the loop. 

Our representation captures sequencing, branching, and looping. 
Sequencing is implicitly captured by the production rule of $\langle G \rangle$, which connects a sequence of $\langle BG \rangle$. 
Branching is captured using $\langle BR \rangle$ while looping is represented using $\langle AL \rangle$ and $\langle GL \rangle$. 
For any given CFG, we can extract the above representation in the following steps. First, we extract all loops in the CFG. Second, we recognize all remaining conditional branches that are not part of a loop's control flow. Finally, using a recursive descent parser on the CFG, we construct $\langle G \rangle$ from the CFG  starting from the entry. 
In our implementation for LLVM IR, we use information provided by LLVM's LoopInfo pass about loops and branches from the CFG to determine whether $\langle BG\rangle$ is either $\langle BB\rangle$, $\langle BR\rangle$, $\langle AL\rangle$ or $\langle GL\rangle$. 

\subsection{Syntax-Guided Execution Count Computation} \label{sec:semantics}

For a program in our representation, we now describe how to obtain symbolic expressions for each basic block to form a symbolic profile.
The process is carried out in two steps: in the first, expressions are built containing \textit{counting functions}.
These counting functions represent number of loop iterations, or the taken/non-taken ratio of branches.
Each counting function is associated with a particular instance of a construct in our representation.
In the second step, we solve for these counting functions to replace them with symbolic expressions of program parameters.

The complete rules for building symbolic expressions for each construct in our representation are shown in Fig.~\ref{fig:semantics}. 
We primarily use the graphical version of our representation (Fig.~\ref{fig:graph-visual}) to describe these rules.

\begin{figure} %
\centering
\footnotesize
\setlength{\baselineskip}{2.8\baselineskip}
\myinfrule
  {BG}
  {\{\#(BG)=1\}}
  {Basic}
\rulesep
\myinfrule
  {G:BG}
  {\{\#(BG)=\#(G)\}}
  {Basic-Reduce}
\rulesep
\myinfrule
  {G:BG~G_1}
  {\{\#(BG)=\#(G),\,\#(G_1)=\#(BG)\}}
  {Sequence-Reduce}
\rulesep
\myinfrule
  {BB:stmts}
  {\{\#(stmts)=\#(BB)\}}
  {Stmt-Association}
\rulesep
\myinfrule
  {BR:if~(C)~G_1,G_2}
  {\{\#(G_1)=\#(BR)\cdot trueRatio(BR)\}}
  {Branch-True}
\rulesep
\myinfrule
  {BR:if~(C)~G_1,G_2}
  {\{\#(G_2)=\#(BR)\cdot(1-trueRatio(BR))\}}
  {Branch-False}
\rulesep
\myinfrule
  {AL:from~S~to~E~in~K~for~G_b,G_e}
  {\{\#(G_b)=\#(AL)\cdot loopCount(AL)\}}
  {Afn-Loop-Body}
\rulesep
\myinfrule
  {AL:from~S~to~E~in~K~for~G_b,G_e}
  {\{\#(G_e)=\#(AL)\}}
  {Afn-Loop-Exit}
\rulesep
\myinfrule
  {GL:while~(N)~G_b,G_e}
  {\{\#(G_b)=\#(GL)\cdot loopCount(GL)\}}
  {Gen-Loop-Body}
\rulesep
\myinfrule
  {GL:while~(N)~G_b,G_e}
  {\{\#(G_e)=\#(GL)\}}
  {Gen-Loop-Exit}

\caption{Operational semantics for the graph language.}
\label{fig:semantics}
\Description{A collection of inference rules that propagate symbolic execution counts through the graph language. The basic and sequence rules pass an incoming count to component graphs and statements. Branch rules multiply the branch count by the true ratio or one minus the true ratio. Affine-loop and general-loop rules multiply the loop count by the corresponding iteration count for the body while passing the original loop count to the exit.}
\end{figure}

Our rules contain the counting function $\#()$ to describe the count of a graph construct defined as the execution count of the construct's incoming edge.
As illustrated in Fig.~\ref{fig:graph-visual}, $\langle G\rangle$, $\langle BB\rangle$, and $\langle BR\rangle$ have a unique incoming edge. 
For $\langle AL \rangle$ and $\langle GL\rangle$, the loop header node has two incoming edges, one from outside, another one from its loop body. 
We consider the one from outside as the incoming edge for the loop constructs.

In addition to the $\#()$ function, we define three semantic functions to handle branches and loops: $trueRatio(BR)$, $loopCount(AL)$, and $loopCount(GL)$. 
The function $trueRatio(BR)$ represents the proportion of times that condition $C$ evaluates to true during the entire program execution. 
The functions $loopCount(AL)$ and $loopCount(GL)$ compute the number of iterations of the loop body for affine and general loops. Note that, if a loop uses an early exit to terminate prematurely, then by definition, its $loopCount$ is the taken iterations instead of full iterations. 

The semantic rules describe how block count flows between different graph components. 
In the semantic rules, each graph construct is in the form of $\textbf{G : components}$ except for the (Basic) rule. 
The form $\textbf{G : components}$ contains an upper-level graph $G$ and the components are the relevant lower-level sub-graphs described in the grammar. 
For example, in the rule of $BR$: if ($C$) $G_1$, $G_2$, $BR$ is the upper-level graph of $G_1$ and $G_2$. 
When evaluating the rule for $BR$, we view $G_1$ as a folded graph with no internal structure. 
The second part shows an inner view of this graph construct, which helps determine its count accordingly.

The (Basic) rule is applied to $BG$ when there is no upper-level graph that uses it as its internal building block. 
Since it is the top-level graph, the count for it is 1. 
Rule (Basic-Reduce) represents the case when the $BG$ is the only internal structure of an upper-level graph $G$. Hence $BG$ and $G$ share the same count. Rule (Sequence-Reduce) shows the case when $G$ is a sequence of $BG$ and $G_1$. The flow comes into $G$, then goes to $BG$, finally passes $G_1$ with all three sharing the same count.  

Rule (Branch-True) and rule (Branch-False) for branch node $BR$ 
split the count of $BR$ into counts for the two component graphs $G_1$ and $G_2$ using the semantic function $trueRatio(BR)$.

In rule (Afn-Loop-Body), $\#(AL)$ represents the number of times this affine loop is visited. 
The $loopCount(AL)$ computes the number of iterations for the loop body for each individual visit. As a result, the final count of the loop body $G_b$ is $\#(AL)\cdot loopCount(AL)$. 
In rule (Afn-Loop-Exit), the loop exit is visited the same number of times as the loop is visited since we now haven't considered early exit of loops. 
Rules for the general loop are similar to rules of affine loop.

Our analysis begins from the top-level graph getting 1 as its count.
The count then flows forward into lower-level subgraphs as defined by the semantics. 
Graphs keep expanding while block count flows until reaching the ground case: $\langle BB \rangle$. At the end, each basic block $\langle BB \rangle$ gets a symbolic formula in terms of $trueRatio(BR)$, $loopCount(AL)$, and $loopCount(GL)$.

\begin{figure}[tbp]
    \centering
    \includegraphics[width=0.4\linewidth]{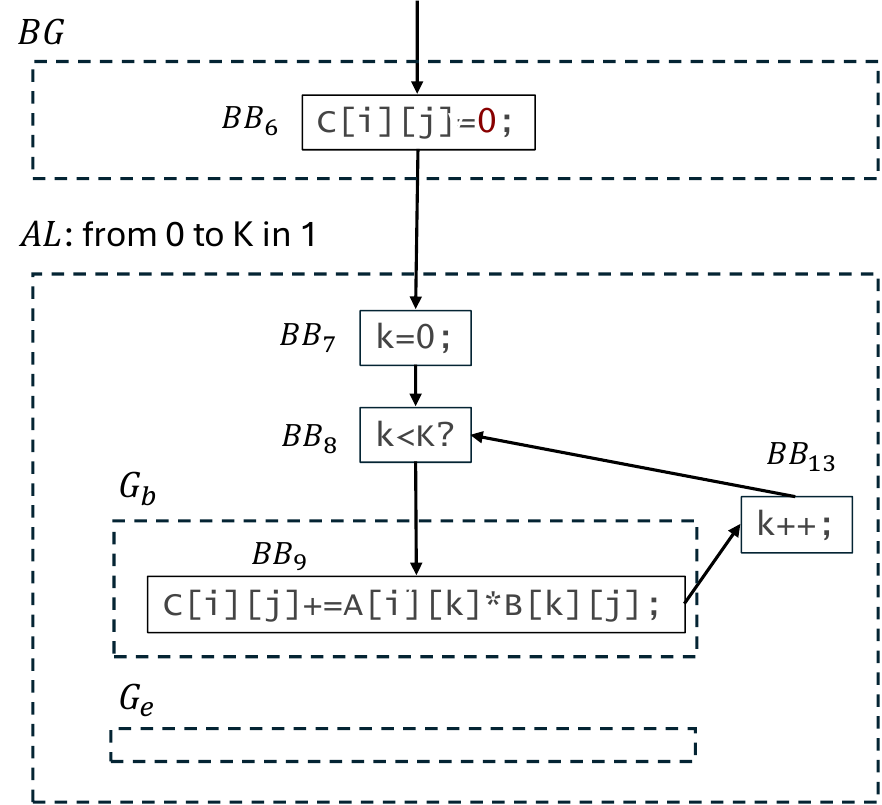}
    \caption{Alignment of partial CFG of Fig.~\ref{fig:naive-matmul-cfg} with the graph representation. Solid boxes are elements in the CFG, dashed boxes show the corresponding graph representation.}
    \label{fig:grammar-example}
    \Description{A portion of the matrix-multiplication control-flow graph aligned with the proposed graph representation. Basic block BB6, which initializes an output element, forms a basic graph. Blocks BB7, BB8, BB9, and BB13 form an affine-loop construct whose body is BB9, containing the multiply-accumulate operation. The affine loop ranges from zero to K in steps of one and has an empty exit subgraph.}
\end{figure}

We apply our representation to a small part of matrix multiply example from Fig.~\ref{fig:naive-matmul-cfg}. We focus on the body of the second nested loop corresponding to lines~\ref{line:cij}-\ref{line:cij-end} of Listing~\ref{lst:naive-matmul}. Fig.~\ref{fig:grammar-example} shows our representation for the CFG. In our representation, $BB_6$ is recognized to be a $BG$, $BB_{7,8,9,13}$ together are recognized to be a $AL$. Among the $AL$, $BB_{7,8,13}$ act as an indivisible part of the loop. The loop body is $BB_9$. Since there are no other instructions after the affine loop, the graph exit $G_e$ is set to be empty. Following rule (Sequence-Reduce), we know that in this case $\#(BG) = \#(AL)$ because they are in the same level and belong to the same upper-level $G$. So we further know that $BB_6$ and the loop have the same count. Diving into the $AL$, $BB_9$ is the loop body. Based on rule (Afn-Loop-Body), the count of the body is the loop iteration count multiplied by the number of times the loop is visited. Since the innermost loop has iteration count $K$, then knowing the loop is visited $MN$ times will give us the result 
\[
\#(BB_9) = \#(AL)\cdot loopCount(AL) = \#(BB_6)\cdot K
\]
The count reflects the result in Fig.~\ref{tab:block-expressions}. The count for $BB_6$ is computed by constructing the complete representation. The counts for $BB_{7,8,13}$ are computed in the same way with the consideration that exiting from loop head will increase the loop head count by one.

\subsection{Early Exits}\label{sec:earlyExit}

Section~\ref{sec:language} and Section~\ref{sec:semantics} show how we can generate a symbolic count for every basic block with the help of several semantic functions. 
We now discuss how we can handle additional early exits. Specifically, we aim to support the three early-exit statements: \texttt{continue}, \texttt{break}, and \texttt{return}, which are treated as normal statements and kept unevaluated in $BB:stmts$ in previous sections. We solve each by adding a small modification to the main framework.

\subsubsection{continue} \label{sec:continue}

\begin{figure}[tbp]
  \centering
  \begin{subfigure}[t]{0.2\textwidth}
    \centering
    \includegraphics[width=0.6\linewidth]{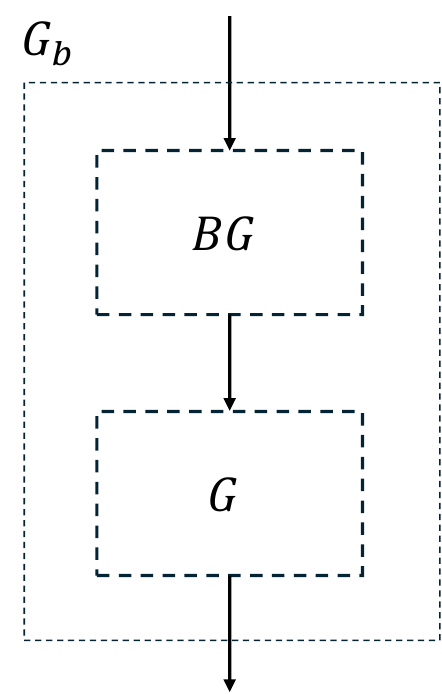}
    \caption{Original}
  \end{subfigure}
  \hspace{0.6cm}
  \begin{subfigure}[t]{0.2\textwidth}
    \centering
    \includegraphics[width=0.6\linewidth]{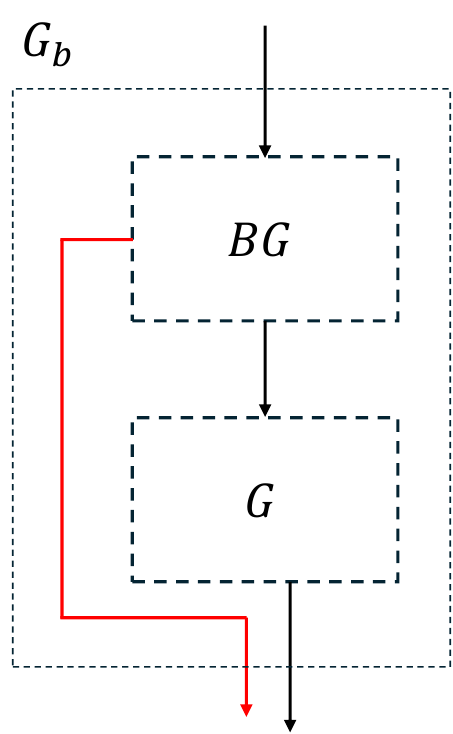}
    \caption{With \texttt{continue}}
    \label{fig:continue-b}
  \end{subfigure}

  \caption{Illustration of flow escapes inside loop body caused by \texttt{continue}. Both (a) and (b) represent a loop body graph $G_b$. (a) shows the original case when the effect of \texttt{continue} is not considered. All flow goes through $BG$ and then goes through $G$. So the count for $BG$ and $G$ in (a) are the same. (b) shows when \texttt{continue}'s effect is considered. Some flow (marked as red) will skip $G$. So the count of $G$ in (b) is less than the count of $BG$.}
  \label{fig:continue-effect}
  \Description{Two loop-body diagrams compare normal sequential control flow with control flow containing a continue statement. In the original graph, all execution flows from the first basic graph into the following graph. In the modified graph, a bypass edge caused by continue exits the loop body before reaching the following graph. Consequently, the following graph receives fewer executions than the first basic graph.}
\end{figure}

The \texttt{continue} statement skips to the end of the loop body. To see the effect of a \texttt{continue}, we go back to the language. As shown in Fig.~\ref{fig:cfg-grammar} and Fig.~\ref{fig:graph-visual}, a loop construct contains a loop body $G_b$, where a \texttt{continue} statement may reside. Since $G_b$ is itself a $G$, it contains a sequence of $BG$. Fig.~\ref{fig:continue-effect} shows the situation when the first $BG$ contains \texttt{continue}. The effect of \texttt{continue} is to cause some flows to exit early before hitting next $G$. The control of some iteration will go directly to the loop end without passing through the second $G$  invalidating the  (Sequence-Reduce) inference rule. 

The corresponding augmented inference rule is shown in Fig.~\ref{fig:continue-augmented}. Here we introduce another semantic function called $escapeRatio()$, which is similar to $trueRatio()$ but measures the percentage of flows that trigger \texttt{continue}. As a result, inside a loop body, the flow can escape whenever a \texttt{continue} appears. The nodes in the following $G$ as shown in Fig.~\ref{fig:continue-b} get less flow. 

\begin{figure}[tbp]
\centering
\small
\myinfrule
      {G:BG ~ G_1}
      {\{\#(BG)=\#(G),\#(G_1)=\#(BG)\cdot(1-escapeRatio(BG))\}} 
      {Sequence-Reduce-Escape}
      \notag
\caption{Operational semantics to handle \texttt{continue}}
\label{fig:continue-augmented}
\Description{An inference rule for propagating counts past a possible continue statement. The first basic graph receives the incoming graph count. The following graph receives that count multiplied by one minus the escape ratio of the first basic graph, accounting for executions that leave the loop body early.}
\end{figure}

The implementation of $escapeRatio(BG)$ can reuse the existing framework. Assuming we are now extracting the $escapeRatio$ for certain $BG_e$, previously introduced framework shows how to extract the count for each basic block inside a graph through the syntax-guided count computation. Hence, treating $BG_e$ as a subgraph enables us to extract count for each internal basic block in it. By setting the incoming edge count of $BG_e$ to be 1, each non-loop basic block inside $BG_e$ gets counts that represent their ratio of being taken. As a result, the sum of all counts for the non-loop basic blocks that contain a \texttt{continue} is exactly the $escapeRatio$ for $BG_e$.

\subsubsection{break}
A \texttt{break} statement can also appear inside a loop and terminates the loop. Our definition of $loopCount$ represents the number of iterations during execution. So the effect of \texttt{break} is captured by the $loopCount$ definition itself.

However, getting a formula that represents the $loopCount$ with \texttt{break} is not trivial. A loop's iteration count with a \texttt{break} requires knowing \textit{when} the \texttt{break} is executed. In the worst case, this would require simulating the loop and even then, the counting functions may not have a closed form. Therefore, although the framework supports such early exit, in our implementation, we do not attempt to get closed form $loopCount$ for loops with \texttt{break}.
Fortunately, none of the ML kernels we study contain \texttt{break}.

\subsubsection{return}
Similar to \texttt{break}, the use of \texttt{return} also complicates our analysis significantly. In the big picture, the effect of \texttt{return} is captured by the definition of $loopCount$ as well. We first distinguish two types of uses of \texttt{return}, one which is inside a loop, and another which is outside a loop. On examination, we found that the use of \texttt{return} in ML kernels is restricted to error detection and handling and normal exit.
That is, machine learning kernels only contain \texttt{return}s that are outside a loop.

\begin{figure}[tbp]
    \centering
    \begin{subfigure}{0.2\textwidth}
        \centering
        \includegraphics[width=0.6\linewidth]{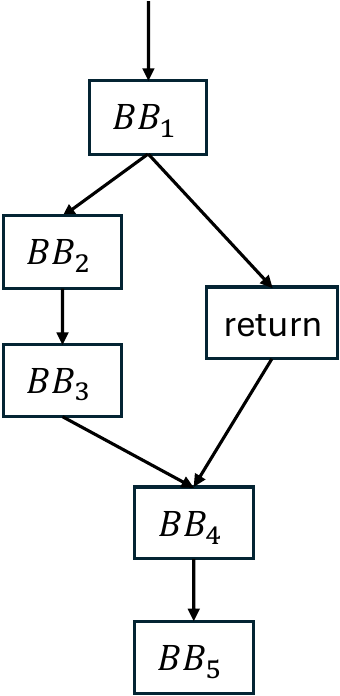}
        \caption{count flow graph}
    \end{subfigure}
    \begin{subfigure}{0.2\textwidth}
        \centering
        \includegraphics[width=0.6\linewidth]{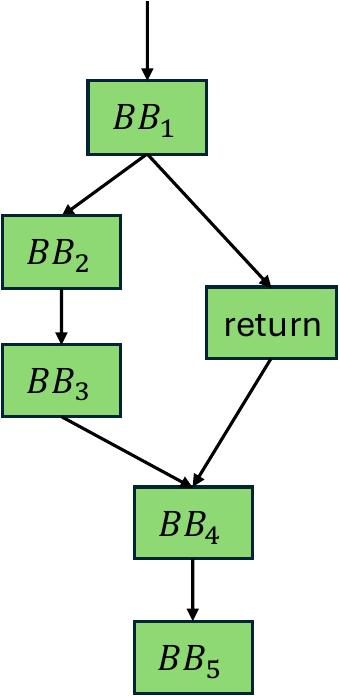}
        \caption{$\#(return) = 0$}
    \end{subfigure}
    \begin{subfigure}{0.2\textwidth}
        \centering
        \includegraphics[width=0.6\linewidth]{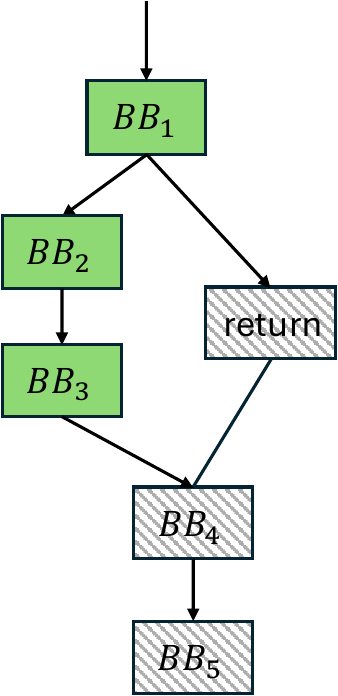}
        \caption{$\#(return)\neq 0$}
    \end{subfigure}
    \caption{This figure shows the topological ordering of basic blocks in an example program. The ordering is based on their count dependency in our framework. Since our framework abstracts away loops and results in a DAG, such topological ordering is always possible. (a) visualizes the ordering $BB_1\to BB_2 \to BB_3 \to $ \texttt{return} $\to BB_4 \to BB_5$. (b) and (c) show two cases when some basic blocks are accepted/invalidated.
    }
    \label{fig:count-flow}
    \Description{Three directed acyclic graphs illustrate count evaluation in the presence of a return. The first graph gives the dependency order from BB1 through BB2 and BB3 and through a return node, with both paths joining at BB4 before BB5. In the second graph, a zero return count leaves every basic-block count valid. In the third graph, a nonzero return count invalidates the return node and its following blocks BB4 and BB5.}
\end{figure}

We handle  \texttt{return}s that happen outside a loop by invalidating the counts of blocks that would not be executed if the \texttt{return} executes.
We first obtain symbolic expressions for all basic blocks in the program without considering the effect of \texttt{return}. 
The counts for each basic block are obtained following their topological order induced by our framework.

Then, during evaluation of symbolic basic block counts, we take the effect of \texttt{return} into consideration. 
The count of each basic block that follows a basic block that contains a \texttt{return} in the topological ordering is invalidated and set to zero, if a block containing \texttt{return} is evaluated to be non-zero.
Fig.~\ref{fig:count-flow} shows an example, where (a) represents the recorded count flow between basic blocks. 
All the basic blocks have a symbolic count associated with them. 
Then, we choose to accept or reject those counts based on the count of \texttt{return} nodes. 
In (b), the \texttt{return} node's count is 0 so it invalidates none of its descendants. 
In (c), the \texttt{return} node is evaluated to be non-zero, so all the following counts are invalidated and set to 0. 

\section{Implementation}\label{sec:impl}

We implement a working symbolic profiling prototype in LLVM. 
Our symbolic profiler works on LLVM IR making it easier to reuse existing analyses as well as support programs written in any language that targets LLVM IR. 
The overall goal of our symbolic profiler is to obtain a symbolic expression for each basic block's execution count.
We break this up into two steps: first, the production of symbolic expressions that contain \textit{counting functions} (e.g. $loopCount()$, $trueRatio()$, $escapeRatio()$), and, second, replacing (or solving) these counting functions with symbolic expressions of only function parameters and constants to yield a symbolic profile.

\begin{figure}[tbp]
    \centering
    \includegraphics[width=0.8\linewidth]{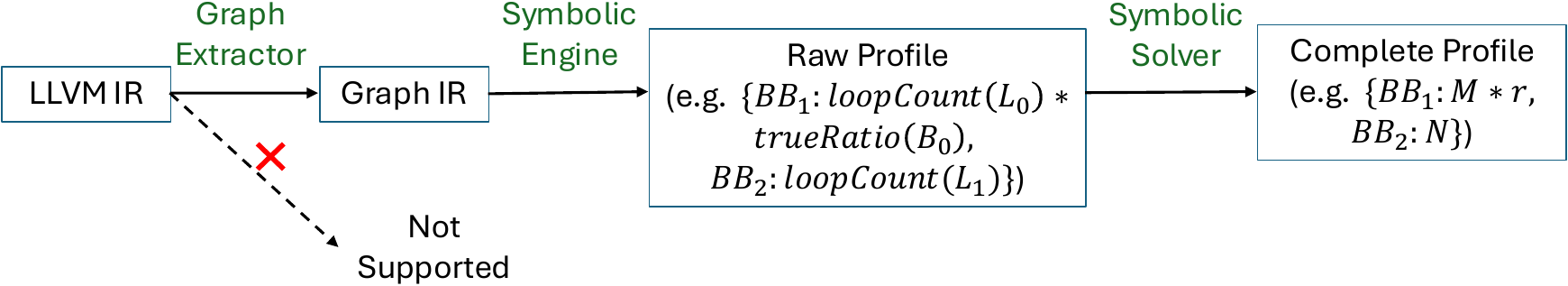}
    \caption{Overall workflow for LLVM IR symbolic profiling.}
    \label{fig:workflow}
    \Description{A left-to-right workflow beginning with LLVM IR. A graph extractor converts supported LLVM control flow into the graph representation, while unsupported structures terminate processing. The symbolic engine constructs a raw profile containing basic-block counts expressed with functions such as loop count and true ratio. A symbolic solver replaces those functions with expressions of program parameters to produce the complete symbolic profile.}
\end{figure}

\subsection{Workflow}\label{sec:workflow}

Symbolic profiling starts from an LLVM IR program (Fig.~\ref{fig:workflow}).
If the input program does not contain unsupported structures such as $return$ inside a loop, then the graph language constructs described in Section~\ref{sec:framework} are extracted from the IR's basic blocks. 
This is done by recognizing all conditional branching and loop structures from the CFG. 
Then the program graph $G$ is constructed by recursively matching CFG's structures with graph structures like $BR$, $AL$. 
A symbolic engine then traverses this extracted graph and propagates counts following the inference rules described in Fig.~\ref{fig:semantics} and Fig.~\ref{fig:continue-augmented}. 
This yields the raw profile which represents execution count of each basic block using counting functions such as $loopCount()$ and $trueRatio()$. 
For the final step, a symbolic solver attempts to solve each counting function to obtain a symbolic expression. 
A complete symbolic profile is obtained when all counting functions are solved. 

\subsection{Symbolic Solving for $trueRatio()$ and $loopCount()$}

We rely on CFG's variable data flow to solve for $trueRatio()$ and $loopCount()$. In our framework, graph-level branches correspond to CFG's conditional branches. Graph-level loops correspond to CFG's loops. Hence both counting functions depend on the value of some CFG variables. Conditional variables determine branches' $trueRatio()$. In the case of loops, LLVM's Scalar Evolution (SCEV) analysis provides trip count of affine loops using variable expressions. The task is now to represent those variables with expressions of program parameters. 

\subsubsection{Simple Case} \label{sec:simple-case}
One type of variables are defined as an arithmetic expression of procedure parameters (usually at the beginning of the procedure). We detect such variables by tracking their data flow for circular dependency. Each variable undergoes a recursive expansion of its operands. A successful expansion leaves us an expression of only procedure parameters. Those expressions have a fixed result when procedure arguments are given meaning that $trueRatio()$ determined by such expressions is always 1 or 0 indicating the corresponding branch will always take the true/false side. Similarly, $loopCount()$ determined by such expressions represents the loop that has the same behavior across the whole procedure execution. Despite their simplicity, counting functions consisting of variables with a fixed expression of procedure parameters are prevalent in ML kernels. 

\subsubsection{Flow-Sensitive Cases} \label{sec:flow-sensitive}
In some cases, variables have flow-sensitive values that show up as definitions through a \texttt{phi} node in LLVM IR with the variable containing different value based on the control flow leading to it. 
Flow-sensitive values complicate solving $loopCount()$.

Listing~\ref{lst:vector-example} illustrates a common occurrence in our benchmarks. 
During vectorization, a compiler will split a single loop into a vectorized part and a remainder part. 
The generated code first checks if the loop is long enough for unrolling in line~\ref{line:vector-branch}. 
If so, the loop is unrolled by the unrolling factor $unr$ and executed in vector mode. 
Finally, the remainder loop handles the iterations that are left over after unrolling. 

In the remainder loop part (Line~\ref{line:remainder-loop}), the $loopCount$ depends on whether the vectorized loop executed. 
If it did, the $loopCount=n\%unr$. 
Otherwise, the remainder loop has $loopCount=n$. 
As a result, solving the remainder loop's $loopCount$ must select which closed-form formula to use ($n\%unr$ or $n$) based on the control flow. 
In this example, determining control flow seems easy since it only involves checking the condition in line~\ref{line:vector-branch}. 
But, in general, tracking control flow can be complicated because it requires tracking all path conditions and their interactions. 

To make things worse, in some cases, multiple control flows are exercised in different proportion during program execution, e.g., imagine the condition in line~\ref{line:vector-branch} is replaced and made dependent on an outer loop index, and for half of the time the loop is unrolled and $loopCount=n\%unr$, while in the other half, $loopCount=n$. Then neither $n\%unr$ nor $n$ can solely represent the $loopCount$. A possible formula will be composite like $\frac{n\%unr ~+~ n}{2}$. 
However, tracking such proportion may require simulating loops, which can be challenging. 
To deal with such flow-sensitive cases and challenges we mentioned above, we introduce \textit{composite expressions}.

\subsubsection{Composite Expressions}

To solve $trueRatio()$ and $loopCount()$ that depend on a symbolic variable whose values depend on control flow, we introduce composite expressions. 
The idea is to create a weighted sum (composite expression) of possible values from different control flows, with weights being the execution proportions of each flow. 
This is possible because we only care about final basic block counts.

For example, suppose that a loop bound is defined by a symbolic
variable \(N\). The total execution count of the loop body is then
\[
N \cdot C_{\mathrm{loop}},
\]
where \(C_{\mathrm{loop}}\) denotes the number of times the loop is
visited. Suppose that \(N\) takes the values \(A\) and \(B\) along two
different execution paths. The total execution count is therefore
\[
A \cdot C_A + B \cdot C_B,
\]
where \(C_A\) and \(C_B\) denote the numbers of times the two paths are
taken. Because
\[
C_{\mathrm{loop}} = C_A + C_B,
\]
the total count can equivalently be written as
\[
C_{\mathrm{loop}}
\left(
A \cdot \frac{C_A}{C_{\mathrm{loop}}}
+
B \cdot \frac{C_B}{C_{\mathrm{loop}}}
\right)
=
C_{\mathrm{loop}}\left(A r_A + B r_B\right),
\]
where
\[
r_A = \frac{C_A}{C_{\mathrm{loop}}}
\qquad\text{and}\qquad
r_B = \frac{C_B}{C_{\mathrm{loop}}}
\]
are the corresponding path-taking ratios. Hence, \(N\) can be replaced
with the composite expression
\[
A r_A + B r_B
\]
without changing the value computed by
$loopCount(\cdot)$. Hence, it is valid to use a composite expression to summarize the changing $loopCount$ because it computes the same final basic block counts as considering each case separately.

\begin{figure}[tbp]
    \centering

    \begin{minipage}[t]{0.48\linewidth}
        \vspace{0pt}
        \begin{lstlisting}[
          style=clean,
          language=C,
          caption={Simplified vectorization code in the TVM add kernel},
          label={lst:vector-example},
          numbers=left,
          numberstyle=\tiny,
          numbersep=6pt,
          xleftmargin=2em,
          framexleftmargin=2em,
          columns=fullflexible,
          breaklines=true,
          basicstyle=\ttfamily\footnotesize,
          escapechar=|
        ]
int i = 0; |\label{line:vector-entry}|
if (n >= unr) { |\label{line:vector-branch}|
    // unrolled body
    for (; i + unr <= n; i += unr) { |\label{line:vector-unroll}|
        body_1
        body_2
        body_3
        body_4
    } |\label{line:vector-unroll-end}|
}
// remainder
for (; i < n; i++) { |\label{line:remainder-loop}|
    body_1
} |\label{line:vector-remainder-end}|
    \end{lstlisting}
    \end{minipage}
    \hfill
    \begin{minipage}[t]{0.42\linewidth}
        \vspace{0pt}
        \centering
        \includegraphics[
            width=\linewidth
        ]{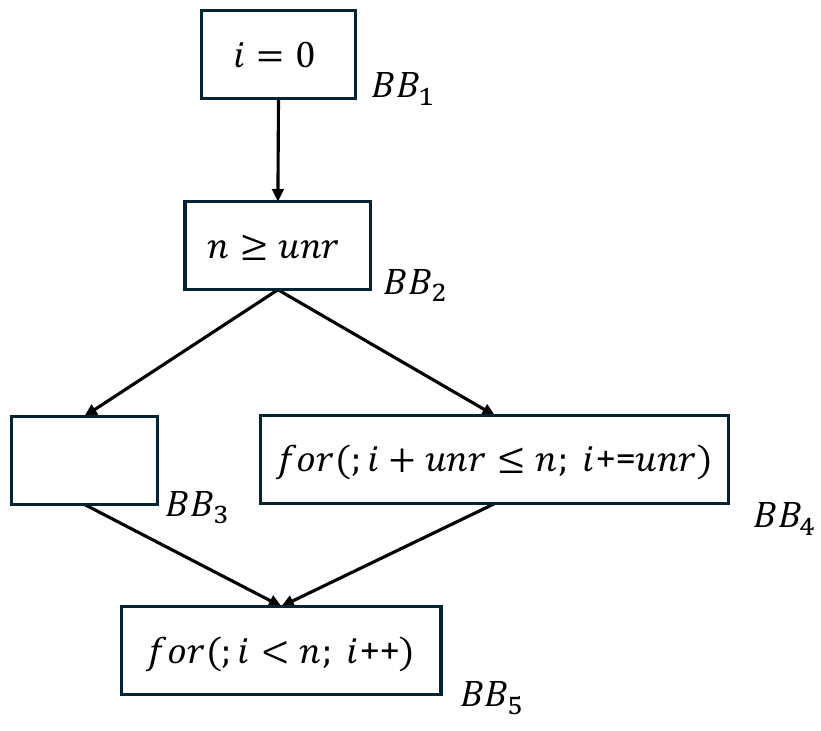}

        \caption{CFG of Listing~\ref{lst:vector-example}. 
        $BB_1$ is the entry, $BB_2$ is the branch, $BB_3$ is
        its empty false path, $BB_4$ is the unrolled loop, and
        $BB_5$ is the remainder loop.}
        \label{fig:vectorization-cfg}
    \end{minipage}
    \Description{The left side contains simplified vectorization pseudocode. It initializes an induction variable, conditionally executes a loop unrolled by factor unr, and then executes a scalar remainder loop. The right side shows the corresponding control-flow graph: BB1 initializes the variable, BB2 tests whether the input is large enough for unrolling, BB3 is the empty false path, BB4 is the unrolled loop, and BB5 is the remainder loop.}
\end{figure}

To capture the execution proportion of each control flow, we reuse the basic block information that is already available in the framework. 
For example, in Fig.~\ref{fig:vectorization-cfg}, the proportion of execution coming from $BB_3 \to BB_5$ (when $n < unr$) and from $BB_4 \to BB_5$ ($n >= unr$) can be represented using their block counts e.g. $\frac{\#(BB_3)}{\#(BB_3) ~+~ \#(BB_4)}$ and $\frac{\#(BB_4)}{\#(BB_3) ~+~ \#(BB_4)}$.

We can then extract the composite $loopCount$ for the remainder loop by reusing the incoming basic block counts $loopCount(BB_5)= \frac{\#(BB_3)*n ~+~ \#(BB_4)*(n\%unr)}{\#(BB_3) + \#(BB_4)}$. Since in our framework the count flows forward, $\#(BB_3)$ and $\#(BB_4)$ are known at $BB_5$. In this way, flow-sensitive cases can be solved based on already available information. The composite technique works for $trueRatio()$ similarly.

To summarize, we first use counting functions e.g. $loopCount()$ and $trueRatio()$ to represent count of each basic block. Later, solving those counting functions takes two forms. The simple form tracks the data flow of the unique CFG variable (or an expression of variables) that determines the counting function to get an expression containing only program arguments, as discussed in Section~\ref{sec:simple-case}. The composite form tracks multiple variables that should be taken based on different control flow separately. We make use of existing basic block count information in the framework to create the composite expression.

\subsection{Solving $trueRatio()$ for Affine Conditions}\label{sec:affine-condition}

If the conditional expression in the branch for which $trueRatio()$ is calculated is loop invariant, we can usually form a symbolic expression containing function parameters using data flow as in the previous section.
However, when branch conditions are not loop invariant, computing a symbolic form of $trueRatio()$ in general is difficult.
We describe how to construct a symbolic form for the most common case in ML kernels, when branch conditions involve a comparison with affine expression over the induction variable. 

\subsubsection{Example} \label{sec:range-example}
We start by showing a representative case when the condition depends on loop's induction variable. 
Listing~\ref{lst:tvm-conv-example} shows the convolution kernel used in image processing machine learning models.
The convolution kernel applies a fixed-size stencil (also known as the kernel) to a larger image.
The stencil accumulates values of neighboring pixels into the output pixel. 
Since images have finite extents, pixels at the boundaries have no neighbors.
The conv kernel must check if a pixel is on the boundary (line~\ref{line:conv-lhc}). Then it performs different operations for internal pixels (at line~\ref{line:conv-A}) and boundary pixels (at line~\ref{line:conv-B}). For our framework to work, we need to compute the $trueRatio()$ for the branch at line~\ref{line:conv-branch}. In other words, we need to know the proportion of pixels that are within the boundary. 

A natural solution is to view the iteration space of the induction variable as a range. While conditions imposed on the induction variable create a refined range that is smaller than the original one. In this example, the original iteration space for $i$ is $ [0,H+1]$ (line ~\ref{line:conv-for}). The conditions define a finer range of $[1,H]$ (line~\ref{line:conv-lhc}). The $trueRatio()$ hence can be represented using fraction of refined range over the original range e.g. $\frac{|[1,H]|}{|[0,H+1]|}$. In the following we aim to generalize and formalize such analysis.

\begin{figure}[tbp]
\centering
\begin{lstlisting}[
  style=clean, language=C,
  caption={Simplified boundary detection code of TVM conv kernel.}, label={lst:tvm-conv-example},
  numbers=left, numberstyle=\tiny, %
  xleftmargin=2em,            %
  framexleftmargin=2em,       %
  columns=fullflexible,       %
  breaklines=true,            %
  basicstyle=\ttfamily\footnotesize,
  escapechar=|
]
for (int i = 0; i< H + 2; i++) { |\label{line:conv-for}|
    bool lower = i != 0, upper = i <= H, cond = lower && upper; |\label{line:conv-lhc}|
    if (cond) |\label{line:conv-branch}|
        A  |\label{line:conv-A}|
    else
        B  |\label{line:conv-B}|
}
\end{lstlisting}
\Description{Simplified convolution boundary-detection pseudocode. A loop iterates from zero through H plus one. Two Boolean tests determine whether the induction variable is nonzero and no greater than H. Their conjunction selects operation A for interior positions and operation B for boundary positions.}
\end{figure}

\subsubsection{Intuition: Counting under Disjunctive Conditions}
As shown in the above example, programmers add conditions over induction variable. Hence the conditions distinguish loop iterations by whether the conditions are satisfied. The goal of the analysis is to automatically compute the iteration counts when the conditions are satisfied. Then we can express the $trueRatio()$ by dividing the satisfied count by total iteration counts.

If the condition can be expressed with disjunctive normal form (DNF), then it naturally becomes a union set problem. In conditions, the DNF can be constructed by grouping logical expressions that are connected with logical \texttt{and}, which are then connected with logical \texttt{or}. 
An expression and-ed together (a conjunctive term) defines a set of induction variable values that satisfy it. 
All conjunctive terms together define the set of all satisfying induction variable values. 
Counting the number of satisfying iterations is equivalent to count the distinct values included by all the sets.

The general approach to solve the union set problem is through the inclusion-exclusion principle. Let $I_i$ be the conjunctive term. Then for two-set case, we have $
|I_1 \cup I_2| = |I_1| + |I_2| - |I_1 \cap I_2|$. For three sets, we have:
$
|I_1 \cup I_2 \cup I_3|
= |I_1| + |I_2| + |I_3| - |I_1 \cap I_2| - |I_1 \cap I_3| - |I_2 \cap I_3|  + |I_1 \cap I_2 \cap I_3|.
$
The number of terms on the RHS grows rapidly with the number of sets that are overlapping making the analysis to support deeper overlap more expensive. Considering both efficiency and the simplicity of conditions we observe in ML kernels, we intentionally limit possible overlapping to be pairwise allowing a distinct element to show up in at most two sets. In our context, it means at any iteration, the induction variable value will not satisfy more than two conjunctive conditions. From our experiments, we only observe single level boundary condition on induction variable. %

In the following, we describe a lightweight analysis that uses the pairwise inclusion-exclusion principle to handle ranges defined through basic comparison and equality relations. 

\subsubsection{Induction Variable Range Analysis} 

We first define a range as a three-element tuple. 
\[
R := \mathrm{range}(s,e,D) \mid \emptyset
\]
\[
\mathrm{range}(s,e,D) :=
\{\,n\in\mathbb{Z}\;\mid\; s\le n\le e,\ n\notin D\,\}
\]

where $s$ and $e$ represent the bounds and $D$ is a set of deleted points. To mimic the analysis in Section~\ref{sec:range-example}, we create the original range $R_0$ which represents the iteration space of the induction variable $i$:
\[
R_0 = \mathrm{range}(S,E,\emptyset),
\]
where $S$ and $E$ are initialized by the corresponding loop bounds. 

Then we define a set of possible conditions that can constrain the value of the induction variable.
Fig.~\ref{fig:cfg-ac} describes the grammar of the conditional expressions we consider. The smallest condition that affects the range of the induction variable is range condition denoted as $\langle RC \rangle$. Here $i$ represents the induction variable, and $v$ represents variable or an expression of variables as described in Section~\ref{sec:simple-case}. We assume $v$ does not depend on $i$. In principle, simple expressions containing $i$ can be manipulated to place $i$ on one side and others on the other, for example $i+1<3$ can be converted to $i<2$. However, in general it is difficult, for example, solving expressions like $i==(i+1)^2$ is not straightforward. We believe that in ML kernels, conditions on induction variables are simple and can be captured by the existing grammar. 

The final affine conditions denoted as $\langle AC \rangle$ is constructed by grouping $\langle RC \rangle$ in disjunctive normal form (DNF). The conjunctive term $\langle T \rangle$ is first constructed by connecting multiple range conditions with logical $\mathbf{and}$. Then the final affine conditions $\langle AC\rangle$ is constructed by connecting $\langle T \rangle$ with logical $\mathbf{or}$.

\begin{figure}[tbp]
\centering
\small
\begin{minipage}{0.95\linewidth}
\[
\begin{array}{rcl}
\langle AC \rangle & ::= & \langle T \rangle 
                     \mid \langle T \rangle\ \mathbf{or}\ \langle AC \rangle \\
\langle T \rangle  & ::= & \langle RC \rangle
                     \mid \langle RC \rangle\ \mathbf{and}\ \langle T \rangle \\
\langle RC \rangle & ::= & i\ \langle rop \rangle\  v 
                     \mid v\ \langle rop \rangle\ i \\
\langle rop \rangle & ::= & \mathtt{<}\ \mid\ \mathtt{>}\ \mid\ \mathtt{<=}\ \mid\ \mathtt{>=}\ \mid\ \mathtt{==}\ \mid\ \mathtt{!=}
\end{array}
\]
\end{minipage}
\caption{Context-Free Grammar for $\langle AC \rangle$}
\label{fig:cfg-ac}
\Description{A context-free grammar for affine conditions in disjunctive normal form. An affine condition is one or more conjunctive terms joined by logical or. Each term is one or more range conditions joined by logical and. A range condition compares induction variable i with an expression v using less than, greater than, less than or equal, greater than or equal, equal, or not equal.}
\end{figure}

Our analysis is to gradually refine the original range $R_0$ based on those conditions. We start by defining the effect of the finest component $\langle RC \rangle$ on ranges through a transfer function $F$. It takes in a range, and returns a refined range based on the given range condition. 
\[
F : R \to R
\]
We show the six transfer functions for the six relational operations in the following. For brevity, we omit $\mathrm{range}(s, e, D)$ on the LHS and just use $R$:
\[
\begin{array}{rcl}
F_{\,i \le v}(R)   &=& \mathrm{range}(s,\min(e,v),\,D) \\
F_{\,i < v}(R)     &=& \mathrm{range}(s,\min(e,v-1),\,D) \\
F_{\,i \ge v}(R)   &=& \mathrm{range}(\max(s,v),\,e,\,D) \\
F_{\,i > v}(R)     &=& \mathrm{range}(\max(s,v+1),\,e,\,D) \\
F_{\,i \neq v}(R)  &=& \mathrm{range}(s,e,D\cup\{v\}) \\
F_{\,i == v}(R)    &=& \begin{cases}
   \mathrm{range}(v,v,D) & s\le v\le e,\\
   \mathrm{\emptyset} & \text{otherwise}.
 \end{cases} 
\end{array}
\]

The effect of the conjunctive term $\langle T \rangle$ can then be derived as follows from the composition of the above transfer functions:

\[
 T = RC_1 \ \mathbf{and}\ RC_2 \ \mathbf{and}\ \cdots \ \mathbf{and}\ RC_m,
\]
\[
F_{T}(R)
  = \big( F_{RC_m} \circ \cdots \circ F_{RC_2} \circ F_{RC_1} \big)(R)
\]

Finally, the effect of affine conditions $\langle AC\rangle$ is to combine ranges that are extracted by conjunctive terms $\langle T\rangle$. We use $I$ to represent range defined by individual $\langle T\rangle$. 
\[
    AC = T_1 \ \mathbf{or} \ T_2 \ \mathbf{or} \ \cdots \ \mathbf{or} \ T_n
\]
\[
I_i := F_{T_i}(R_0) = \operatorname{range}(s_i,e_i,D_i), \quad i=1,\dots,n.
\]
Going to the end, we need to count the number of elements in the original range and that in the refined ranges. We introduce a helper function $\Omega$ to handle range overlapping so that no element is counted more than once in the refined ranges.
\[
\Omega : R \times R \to R
\]
\[
\Omega(I_i, I_j)\ = \ \operatorname{range}\bigl(\max(s_i, s_j),\, \min(e_i, e_j),\, D_i \cup D_j)
\]
We can then get the set $\mathcal{I}$ of all ranges defined by conjunctive terms, and the set $\mathcal{O}$ of pairwise overlapping.
\[
\begin{array}{rcl}
\mathcal{I} &:=& \{I_i \mid i=1,\dots,n \,\},\\
\mathcal{O} &:=& \{\, \Omega(I_i,I_j) \ \mid 1 \le i < j \le n \,\}.
\end{array}
\]
The last piece is the $len$ function to compute the number of elements within a range for the case when iteration step size is 1. Support for different step size requires a small change to the $len$ function which we omit here for simplicity.
\[
len(range(s,e,D)) = \begin{cases}
    (e-s+1) -|D\cap[s,e]| & s\le e, \\
    0 & \mathrm{otherwise}\\
\end{cases}
\]

Finally, the $trueRatio$ of a set of affine conditions ($AC$) can be computed by counting the number of elements in the original range and that satisfy the affine conditions. 
\[
trueRatio_{AC} = \frac{\sum_{I\in \mathcal{I}} len(I) - \sum_{O \in \mathcal{O}} len(O)}{len(R_0)} 
\]
This analysis enables us to extract precise $trueRatio()$ for induction-variable-dependent branches inside loops. 

\subsubsection{Example}

For illustration purposes, we apply the introduced analysis to the conv example in Section~\ref{sec:range-example}. First we have the original range defined in line~\ref{line:conv-for}:
\[
    R_0 = range(0,H+1,\emptyset)
\]
The affine condition is defined by a single conjunctive form that contains two range conditions:
\[
    AC = T= (i \neq 0) \ \mathbf{and}\ (i \le H) 
\]
By applying the transfer function $F$, we can get the refined range:
\begin{align} 
    F_{(i \neq 0 \ \textbf{and}\ i \le H)} (R_0) &= (F_{i \le H} \circ F_{i\neq 0})(R_0) \notag\\ 
    &= F_{i \le H} (F_{i\neq 0} (R_0)) \notag \\ 
    &= F_{i\le H}(range(0, H+1, \{0\})) \notag\\
    &= range(0, min(H, H+1),\{0\}) \notag\\ 
    &= range(0,H,\{0\}) \notag
\end{align}
Since there is only one conjunctive term in the affine conditions, there is no overlapping range which simplifies the computation of the final $trueRatio$. Based on the $trueRatio$ formula and the $len$ function, we get:
\begin{align} 
    trueRatio_{AC} &= \frac{\sum_{I\in \mathcal{I}} len(I) - \sum_{O \in \mathcal{O}} len(O)}{len(R_0)} \notag\\ 
    &= \frac{len(range(0,H,\{0\}))}{len(range(0,H+1,\emptyset))} \notag\\ 
    &= \frac{H-0+1-1}{H+1-0+1} \notag\\ 
    &= \frac{H}{H+2} \notag
\end{align}
Given the above formula, for $H=100$, we will have $trueRatio = \frac{100}{102}$ which can be used in other computations generated by the framework. 

\section{Evaluation}\label{sec:evaluation}

To evaluate our technique, we use the TVM compiler~\cite{TVM} as the source of general machine learning workloads. Compared to other machine learning (ML) frameworks like PyTorch \cite{paszkePyTorchImperativeStyle2019} and TensorFlow \cite{tensorFlow}, the advantage of TVM is that it allows detailed control in its own optimization pipeline.
Combined with the large set of low-level kernel implementations, we therefore have access to a broad space of optimized kernels. 
In the following sections, we show results of experiments that evaluate: 1.~the applicability of our framework to these machine learning kernels, 2.~the execution time of instrumentation compared to our technique.

\subsection{Evaluation Methodology}
\subsubsection{Machine Learning Operators}
Machine learning workloads can be viewed as a data flow graph whose nodes are ML operators and edges represent flow of tensors (i.e., inputs and outputs).
For our evaluation, we build a representative set of ML operators by characterizing 50 ML models from the ONNX zoo \cite{onnx-model-zoo}.
These models perform tasks ranging from vision to language processing. 
We find in total 78 types of operators in these models. The full list is shown in Table~\ref{tab:operator-list}.

\begin{table}[t]
  \caption{The 78 operator types grouped by functionality.}
  \centering
  \scriptsize
  \setlength{\tabcolsep}{3pt}
  \renewcommand{\arraystretch}{0.92}

  \begin{tabularx}{\linewidth}{
    @{}
    >{\raggedright\arraybackslash}p{0.21\linewidth}
    >{\raggedright\arraybackslash}X
    @{}
  }
    \toprule
    \textbf{Category} & \textbf{Operators} \\
    \midrule

    \textsc{NN layers}
    &
    \texttt{matmul, gemm, conv, conv\_transpose, batch\_norm,
    instance\_norm, lrn, lstm, prelu, average\_pool,
    global\_average\_pool, maxpool, softmax}
    \\

    \textsc{Elementwise unary}
    &
    \texttt{abs, not, sqrt, reciprocal, exp, log,
    erf, floor, neg, relu, leaky\_relu, tanh, sigmoid,
    hardmax, clip, ceil, round}
    \\

    \textsc{Elementwise binary}
    &
    \texttt{add, sub, mul, pow, div, and, less, greater, less\_or\_equal}
    \\

    \textsc{Tensor reshaping}
    &
    \texttt{cast, reshape, squeeze, unsqueeze, flatten, expand,
    transpose, concat, split, slice, tile, pad, compress, resize}
    \\

    \textsc{Reduction}
    &
    \texttt{argmax, reduce\_max, reduce\_mean, reduce\_min,
    reduce\_sum, sum, max, nonzero, topk, nms, cumsum}
    \\

    \textsc{Utility \& others}
    &
    \texttt{where, shape, upsample, roi\_align, identity, constant,
    constant\_of\_shape, category\_mapper, dropout, range,
    loop, scan, gather, scatter}
    \\

    \bottomrule
  \end{tabularx}

  \label{tab:operator-list}
\end{table}

Operators describe the intended computation task. A kernel reflects the real implementation of an operator. For example, an ML model can use multiple matmul operators, and if the input sizes for these operators are different, they will result in different matmul kernels optimized for different input sizes. In our evaluation, for each operator, we investigate its general kernel implementation that admits any sized inputs. In the remaining part, we will simply use the term kernel. 

We evaluate all the shown 78 kernels in TVM. For the majority, we adopt the implementation from TVM's built-in operator inventory. For operators not found in the inventory like LSTM and RoiAlign, we manually construct those through TVM's IR according to ONNX 1.22.0 \cite{onnxOperators} definition. 
These operators then go through optimization and lowering using TVM version 0.21.0 following their official optimizing instructions~\cite{tvmTutorialE2E}. 
None of the kernels we evaluate on are input-size specific.

\subsubsection{Baseline}
To the best of our knowledge, only dynamic instrumentation obtains exact counts for basic block executions, therefore we use it as our baseline. Concretely, we use LLVM's PGO instrumentation pass. In the following, we use the basic block profiling results generated by PGO as the ground truth to verify the results generated by our symbolic method. Later, we compare the execution time of the two methods. %

\subsection{Applicability}
In this section we evaluate the applicability of our symbolic method on the 78 ML kernels shown in Table~\ref{tab:operator-list}.  

\subsubsection{Kernel Complexity}
Table~\ref{tab:llvm-complexity} characterizes the top 42 kernels ranked by the number of total basic blocks. 
As an attempt to measure complexity, we use ten basic CFG metrics, the number of basic blocks \text{BB}, the number of $\phi$-functions \text{Phi}, the maximum loop nesting depth \text{MLD}. A higher number of basic blocks represents larger kernels. 
A higher number of phi nodes and a deeper loop depth capture branching and loop complexity.

The other five metrics: number of $trueRatio()$ counting functions \text{TR}, $loopCount()$ functions \text{LC}, early exits \text{EE}, composite expressions \text{CE}, and affine conditions \text{AC} capture the analysis complexity for our framework. 
The number of required counting functions $trueRatio$ and $loopCount$ shows how variant the block count formula can be. 
In addition, the presence of early exits, flow-sensitive behaviors, and induction-variable-dependent conditions make our symbolic analysis more challenging. At last, there are two metrics of unsupported cases: the data dependent behaviors \text{DD} and non-affine conditions \text{NAC}. 

Among all the kernels, there exist both very simple kernels and the most complicated ones. The \texttt{batch\_norm} kernel contains 51 loops and 55 conditional branches. Inside those loops, there are 13 places where early exits happen and the loop is terminated prematurely. Additionally, there are also 11 cases when the loop/branch has different behavior according to the control flow.

\subsubsection{Experiment}
After extracting the 78 kernels as LLVM IR, we run both LLVM's PGO instrumentation pass and our symbolic method on each of the kernels. 
From PGO we get the concrete basic block profiles by providing concrete inputs. 
Our symbolic tool returns a symbolic profile that assigns each basic block with a symbolic expression. 
We currently use SMT-LIB to record those expressions. 
We then substitute input symbols with the same values we have given to PGO to instantiate the symbolic result. We empirically verify the correctness of our symbolic expressions by matching the evaluation result against the PGO result over sizes from 64 to 8192.

Not all symbols in our symbolic expressions map directly to program inputs.
In the TVM kernels, we find unsolvable symbols (i.e. which cannot be reduced to an expression containing only program inputs) in many symbolic expressions.
In the kernels we look at, these symbols correspond to return values of procedure calls in the kernel's body. 
Most commonly, these are calls to the TVM memory allocator. 
The values of these symbols do affect basic block counts because when the procedure call fails, the control exits the function.
For now, we assume all such calls succeed and assign a constant value 1 to those symbols to reflect the normal flow, although we could also allow users to override these values. There are also calls to utility function \texttt{max}. We handle these by replacing the corresponding symbols with formulae that implement the corresponding functionality, i.e. \texttt{max(a,b)} is converted to $ITE(a>b, a, b)$.

\subsubsection{Result}
We successfully support 73 out of 78 kernels where we extract correct symbolic expression \emph{for every basic block}. The remaining five kernels are \texttt{pad}, \texttt{prelu}, \texttt{topk}, \texttt{cumsum} and \texttt{loop}, two of which are partially supported, and three exceed the scope of symbolic profiling. 

The two partially supported kernels are \texttt{pad} and \texttt{prelu} as shown in Table~\ref{tab:llvm-complexity}. 
The \texttt{pad} kernel contains a non-affine condition where the induction variable carries out bit-wise or operation optimizing an induction variable increment. Such a non-affine condition falls out of the scope we define in Section~\ref{sec:affine-condition}. However, we still solve 18 out of 19 basic block counts for \texttt{pad}. In \texttt{prelu}, computation is only performed for elements that are greater than 0 which is a data-dependent behavior. 
In this case we solve 11 out of 14 basic blocks. 

In the remaining three kernels, \texttt{loop} is a ONNX graph-level construct that repeats input kernels, hence it is not a real kernel. The implementation of \texttt{topk} contains a data-dependent sorting algorithm, which is beyond the scope of static symbolic profiling. 
The \texttt{cumsum} kernel in TVM is a wrapper that hands over the computation to the corresponding TVM runtime routine.

\begin{table} %
\caption{Static complexity metrics extracted from LLVM IR kernels. The entries represent the number of instances in that category. 
BB: basic blocks; Phi: phi-nodes; TR: symbolic $trueRatio$; LC: symbolic $loopCount$; 
MLD: max loop depth; EE: early exits; CE: composite expressions as discussed in Section~\ref{sec:flow-sensitive}; AC: affine conditions as discussed in Section~\ref{sec:affine-condition}.; DD: data dependent blocks; NAC: non-affine conditions over induction variable. The highlighted kernels contain unsolvable block counts.}
\centering
\small
\setlength{\tabcolsep}{4pt}
\renewcommand{\arraystretch}{0.77}
\begin{tabular}{lrrrrrrrrrr}
\toprule
\textbf{Name}&\textbf{BB}&\textbf{Phi}&\textbf{TR}&\textbf{LC}&\textbf{MLD}&\textbf{EE}&\textbf{CE}&\textbf{AC}&\textbf{DD}&\textbf{NAC}\\
\midrule
lstm&639&348&208&267&3&0&1&0&0&0\\
batch\_norm&147&79&110&51&4&13&11&0&0&0\\
lrn&59&29&42&20&4&3&3&0&0&0\\
conv\_transpose&55&23&42&14&5&0&1&3&0&0\\
conv&41&25&26&11&7&0&1&3&0&0\\
softmax&39&31&30&12&2&2&2&0&0&0\\
average\_pool&30&14&16&10&4&2&2&0&0&0\\
global\_average\_pool&27&21&18&9&4&1&1&0&0&0\\
instance\_norm&27&13&16&9&4&1&1&0&0&0\\
hardmax&20&16&16&5&2&1&1&0&0&0\\
\textcolor{red}{pad}&19&7&20&2&2&0&1&3&0&\textcolor{red}{1}\\
reduce\_mean&19&15&16&5&2&1&1&0&0&0\\
maxpool&15&6&8&5&4&1&1&0&0&0\\
and&14&6&12&4&2&2&2&0&0&0\\
nms&14&13&12&4&1&1&1&0&0&0\\
not&14&6&12&4&2&2&2&0&0&0\\
\textcolor{red}{prelu}&14&7&12&2&2&0&0&0&\textcolor{red}{3}&0\\
argmax&13&11&12&2&2&0&0&0&0&0\\
gemm&12&12&8&4&3&0&0&0&0&0\\
matmul&12&12&8&4&3&0&0&0&0&0\\
nonzero&11&5&8&3&2&1&1&0&0&0\\
transpose&11&7&10&4&2&1&2&0&0&0\\
reduce\_max&10&11&8&3&2&0&0&0&0&0\\
reduce\_min&10&11&8&3&2&0&0&0&0&0\\
reduce\_sum&10&11&8&3&2&0&0&0&0&0\\
sum&10&11&8&3&2&0&0&0&0&0\\
abs&9&4&6&3&2&1&1&0&0&0\\
add&9&4&6&3&2&1&1&0&0&0\\
cast&9&4&6&3&2&1&1&0&0&0\\
ceil&9&4&6&3&2&1&1&0&0&0\\
clip&9&4&6&3&2&1&1&0&0&0\\
constant\_of\_shape&9&4&6&3&2&1&1&0&0&0\\
div&9&4&6&3&2&1&1&0&0&0\\
dropout&9&4&6&3&2&1&1&0&0&0\\
erf&9&4&6&3&2&1&1&0&0&0\\
exp&9&4&6&3&2&1&1&0&0&0\\
floor&9&4&6&3&2&1&1&0&0&0\\
greater&9&4&6&3&2&1&1&0&0&0\\
leaky\_relu&9&4&6&3&2&1&1&0&0&0\\
less&9&4&6&3&2&1&1&0&0&0\\
less\_or\_equal&9&4&6&3&2&1&1&0&0&0\\
log&9&4&6&3&2&1&1&0&0&0\\
\bottomrule
\end{tabular}
\label{tab:llvm-complexity}
\end{table}

\subsection{Analysis and Execution Time}
We now compare the time for analyzing programs and obtaining profiles using a traditional instrumentation method (PGO) and our symbolic method.

\subsubsection{Experiment Setup}
For PGO, we use LLVM's PGO which uses the Knuth-Stevenson algorithm~\cite{knuthOptimalMeasurementPoints1973} for counter placement. 
The LLVM IR is compiled to generate the instrumented version program which is later executed to generate exact profile for specific inputs. 
In our symbolic method, the LLVM IR is statically analyzed to obtain a set of symbolic formulae.

The analysis time is the time spent on compilation and static analysis.  
For PGO, this is the compile time for the LLVM IR with instrumentation pass enabled.
The PGO compilation accepts LLVM IR and emits LLVM IR. 
For our symbolic method, the analysis overhead is the time spent on static analysis to extract all the symbolic formulae. 

The execution time is the time  required to generate the profile after a program input is given. For PGO, this is the time taken to run the instrumented program on that input.
For our method, it is the time to evaluate the symbolic formulae with program inputs. We have implemented two evaluation processes. The first one uses Z3~\cite{demouraZ3EfficientSMT2008} for value substitution and simplification. The second one uses LLVM to compile the SMT-LIB symbolic formulae into a binary, eliminating the expensive look up and substitution process. 
We show the time for both. 

We conducted the experiments on a machine with an Intel i7-12700H CPU and 32\,GB of RAM. 
For each benchmark kernel, we set the input matrices to have square shapes. 
The widths of the matrices are 64, 128, 256, 512, 1024, 2048, 4096, 8192. 
For each input, we run each experiment 3 times and report the average.

\subsubsection{Analysis Time}
\begin{figure} %
    \centering
    \includegraphics[width=0.95\linewidth]{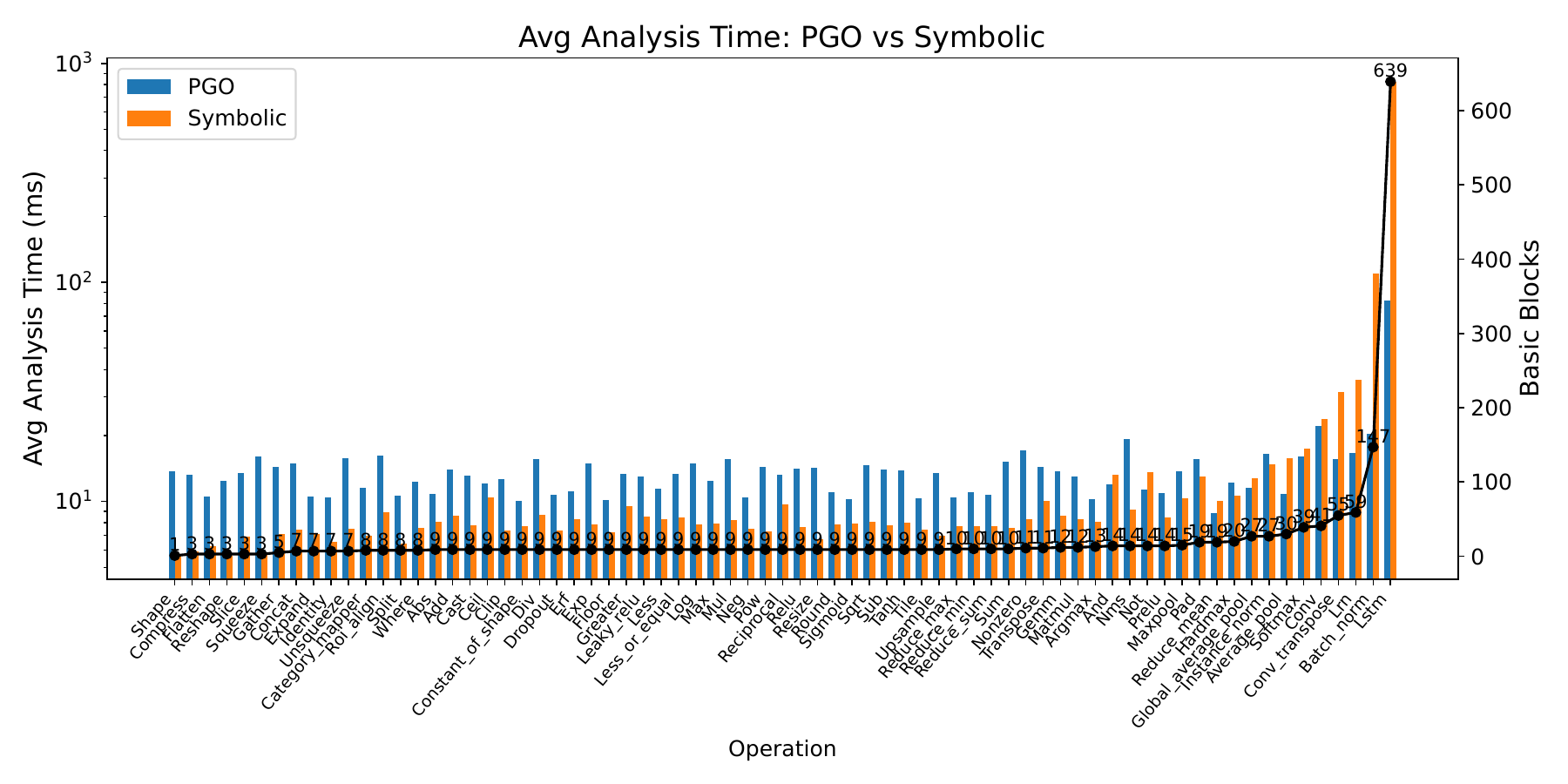}
    \caption{Analysis time of PGO and our symbolic method. The line represents the number of basic blocks in each kernel.}
    \label{fig:all-init}
    \Description{A combined bar and line chart comparing average analysis time across kernel operations. Paired bars give the PGO and symbolic-analysis times on a logarithmic left axis, while a line gives the number of basic blocks on the right axis. The symbolic method is generally faster for small kernels but becomes more expensive as structural complexity increases. Batch normalization and LSTM have the largest basic-block counts and the highest symbolic-analysis times, with LSTM reaching 639 basic blocks.}
\end{figure}

Fig.~\ref{fig:all-init} shows the analysis time for the two methods. For very small kernels with fewer than 20 basic blocks, our method takes less time compared to PGO. However, it is also more  sensitive to the total number of basic blocks in a kernel. For large kernels like \texttt{lstm} and \texttt{batch\_norm}, our method requires more analysis time. This is due to the two-stage approach, where we first create expressions using counting functions, and later we replace them with symbolic formulae of program inputs. This implementation's complexity is currently quadratic over the number of basic blocks and the number of counting functions.

\subsubsection{Execution Time}
\begin{figure} %
    \centering
    \includegraphics[width=\linewidth]{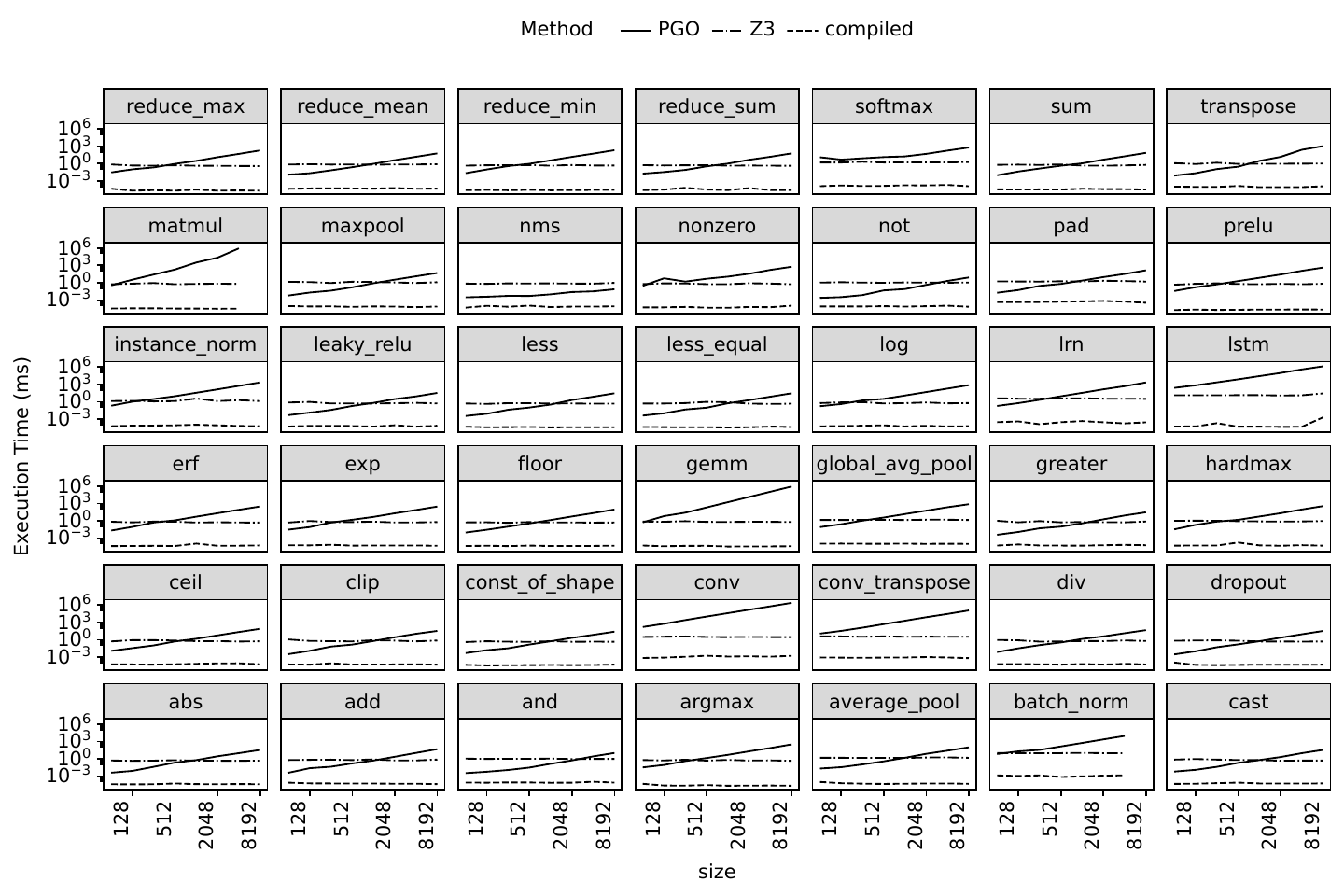}
    \caption{Execution time of PGO and our symbolic method in getting basic block counts for 42 kernels. Each subplot represents the profiling time for a specific kernel with increasing input size. The solid line shows PGO time. And the two dashed lines represent ours.}
    \label{fig:all-exec}
    \Description{A grid of forty-two log-scale plots, one for each kernel. Each plot compares PGO, Z3-based symbolic evaluation, and compiled symbolic evaluation as the input size increases from 128 to 8192. PGO execution time generally rises rapidly with input size, while both symbolic methods remain nearly constant. Compiled symbolic evaluation is consistently the lowest of the three methods.}
\end{figure}

\begin{figure} %
    \centering
    \includegraphics[width=0.5\linewidth]{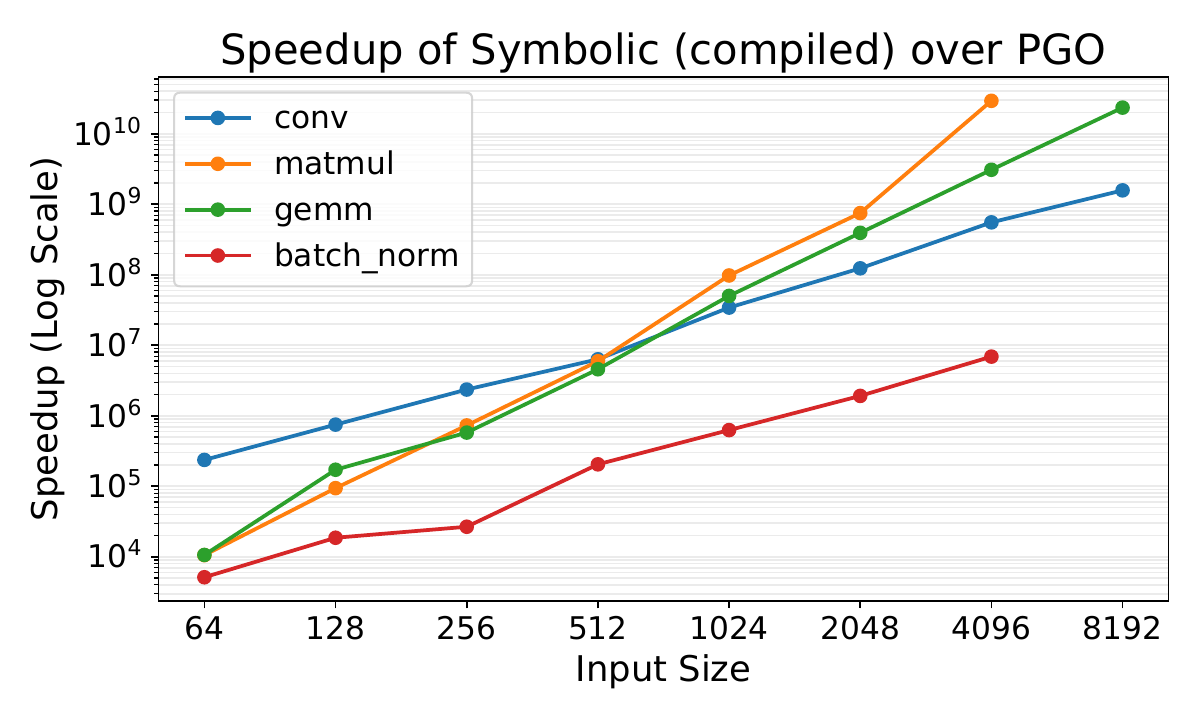}
    \caption{Speedup of Symbolic (compiled) over PGO for computation-heavy kernels.}
    \label{fig:speedup}
    \Description{A logarithmic line chart of compiled symbolic profiling speedup over PGO for convolution, matrix multiplication, general matrix multiplication, and batch normalization. Input size increases from 64 to 8192. Speedup increases with input size for every kernel, ranging from thousands at small sizes to millions or billions at large sizes. Matrix multiplication and general matrix multiplication achieve the largest speedups.}
\end{figure}

Fig.~\ref{fig:all-exec} shows the execution time of two methods for different input sizes. Due to the space limitation, we show the top 42 kernels ranked by the number of total basic blocks (the rest can be found in the appendix in the supplemental material). From the figure we can see, for our symbolic method, the evaluation time (for both Z3 and compiled) is always a constant. While the PGO execution time increases as inputs. Also note that both y-axis and x-axis use log-scale, the gap between two methods soon becomes larger as input size increases. In all the kernels, the compiled version outperforms PGO for all input sizes.   

For computation-heavy kernels like \texttt{matmul} and \texttt{conv}, the symbolic method provides over $10^9\times$ speedup when the input size is large. In Fig.~\ref{fig:speedup}, we plot the speedups for \texttt{batch\_norm}, \texttt{conv}, \texttt{matmul}, and \texttt{gemm}, the speedup increases rapidly with input size.

\subsection{Use in Auto-Tuning}

Basic block profiles are long associated with actual execution, limiting their use to explaining program behavior rather than predicting it.
Our method enables static basic block profiles allowing profiles to have broader application like in program performance modeling. 
In this section, we show how statically available basic block profiles can be used to build a performance model.

We develop a  profile-based cost model for kernel auto-tuning \cite{balaprakashAutotuningHighPerformanceComputing2018, anselOpenTunerExtensibleFramework2014, chen-xgboost-2016}. 
Machine learning kernels often undergo architecture-specific and input-specific optimizations to achieve the best deployed performance. 
Auto-tuning empirically searches for the best optimization in a large tuning space \cite{chen-xgboost-2016, schulzePyATFConstraintBasedAutoTuning2025}. 
Currently, auto-tuning relies on empirical evaluation to guide the search process, making large evaluation overheads the bottleneck. 
By replacing empirical evaluations with cost models, we can reduce auto-tuning's overheads.

For a case study of using kernel profiles to build a performance model to facilitate auto-tuning, we create a simple decision-tree-based model using XGBoost \cite{chen-xgboost-2016}. 
Our model uses a kernel's LLVM IR \textit{instruction map} as its input features.
These are derived from kernel profiles by multiplying basic block counts by the per instruction counts inside that basic block. 
We then evaluate this model in the TVM \cite{TVM} auto tuning framework.

\subsubsection{Experimental Setup}
We use TVM MetaSchedule framework, where a custom cost model can be used to improve auto-tuning of kernel schedules. 
The search strategy uses default `evolutionary' with population size to be 64. 
Thus at each trial, the auto-tuner will propose 64 kernel schedules while the cost model helps decide which one wins and will be evaluated.

We use the `conv2d' kernel from TVM inventory with 512x512 image size and 64 channel size, which is commonly used in today's ML models. 
We compare our model with a random baseline and the default state-of-the-art XGB model~\cite{chen-xgboost-2016} used by TVM internally. 
Both TVM's internal model and our model use XGBoost framework, but ours uses the LLVM IR instruction map as features, while the default uses features extracted from TVM IR.

We run the auto-tuning experiment with budget being 300 trials, which means there can be only 300 empirical evaluations on kernels. 
To account for the effect of random seeds, we ran experiments with 30 different seeds and report the aggregate result.

\subsubsection{Results and Analysis}
\begin{figure} %
    \centering
    \begin{subfigure}[t]{0.48\textwidth}
        \centering
        \includegraphics[width=\linewidth]{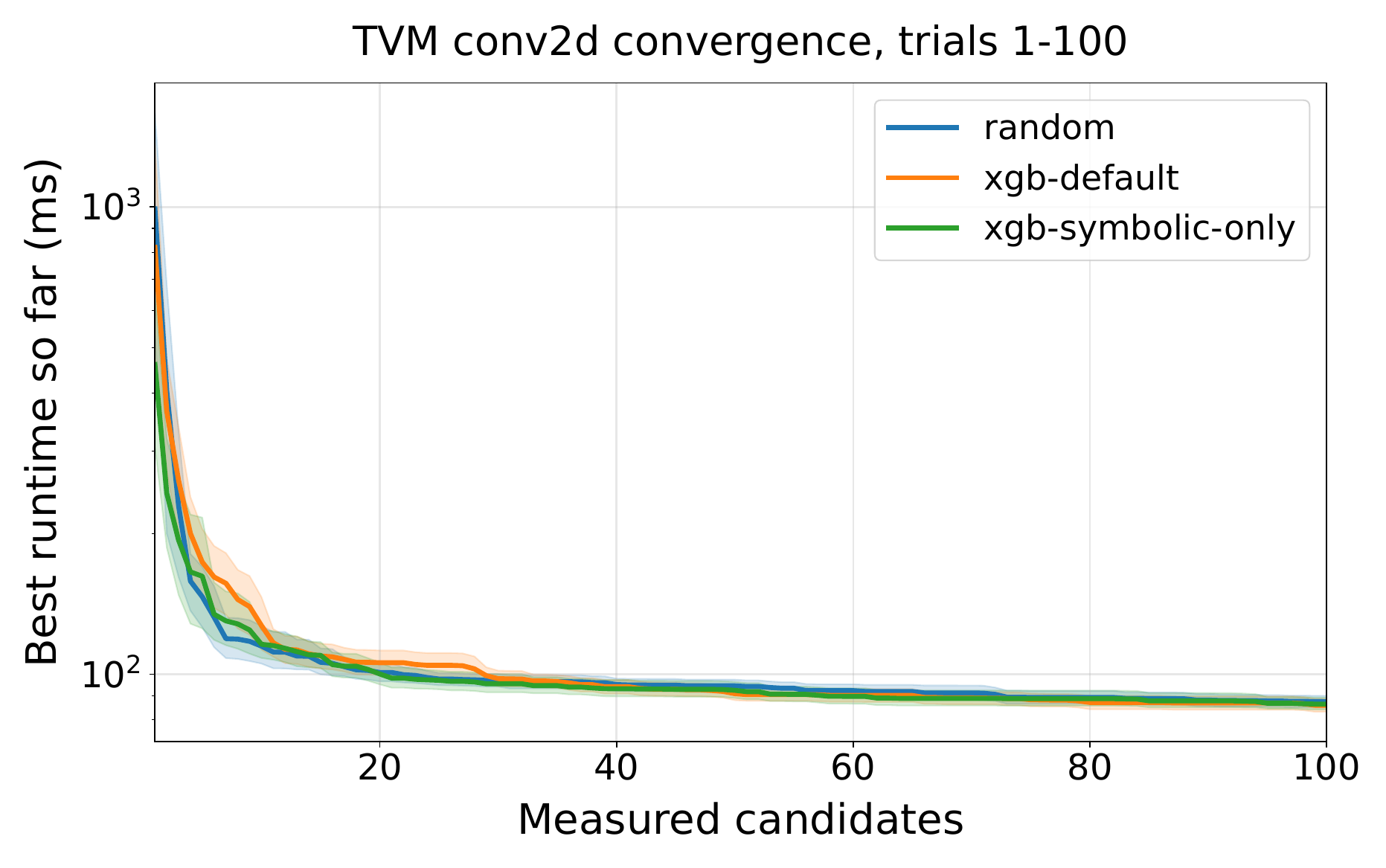}
        \caption{Trials 1 to 100.}
        \label{fig:convergence-1to100}
    \end{subfigure}
    \hfill
    \begin{subfigure}[t]{0.48\textwidth}
        \centering
        \includegraphics[width=\linewidth]{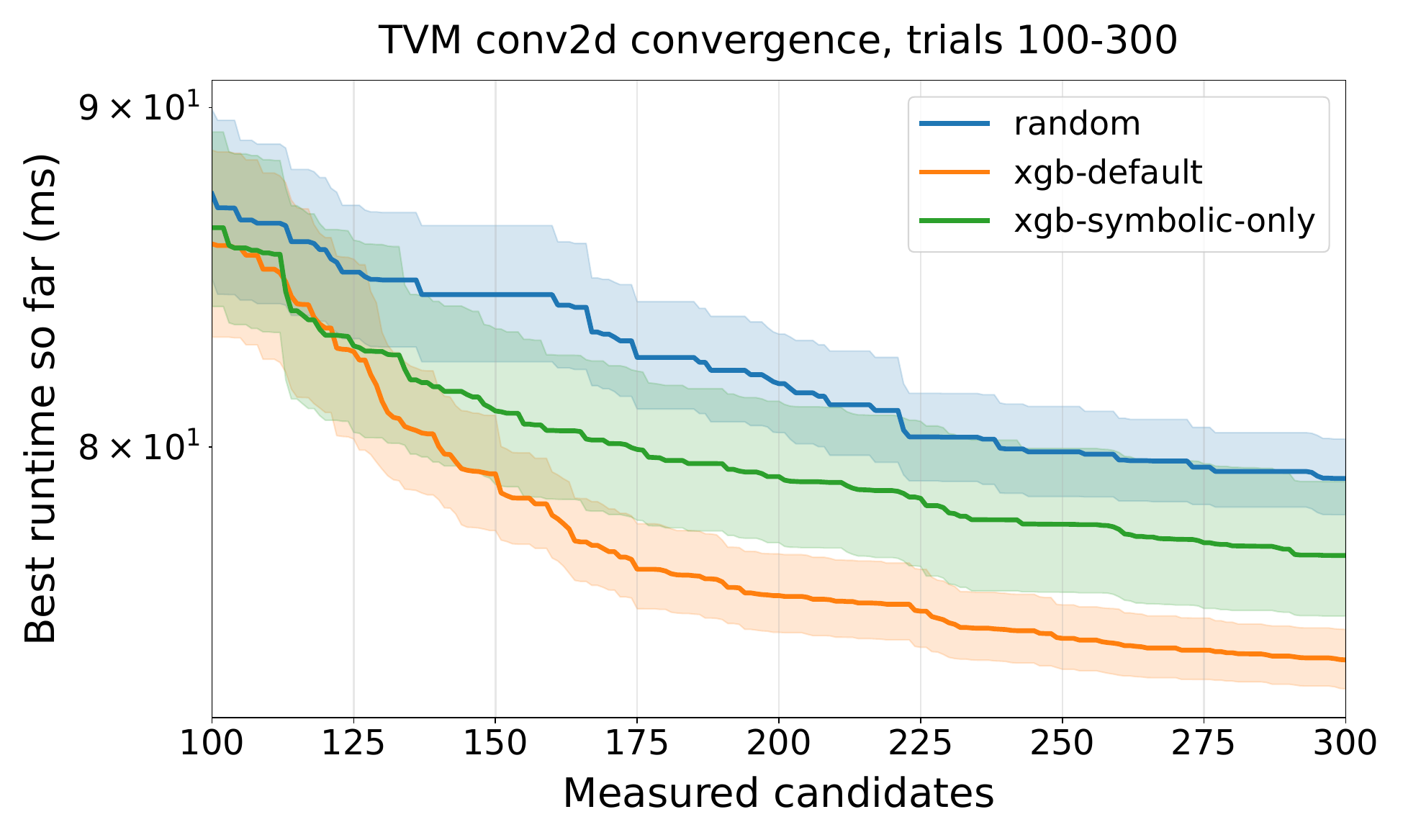}
        \caption{Trials 100 to 300.}
        \label{fig:convergence-100to300}
    \end{subfigure}

    \caption{Convergence of best kernel runtime over trials. Each solid line represents mean of the 30 seeded runs, while the shaded area is the 95\% confidence interval. (a) is the result for first 100 trials. (b) shows the finer-grain runtime changes from 100-300 trials.}
    \label{fig:convergence}
    \Description{Two line charts compare random search, the default XGBoost model, and the symbolic-profile XGBoost model for TVM convolution tuning. During trials 1 through 100, all three methods rapidly reduce the best runtime from several hundred or more milliseconds to below 100 milliseconds and remain close to one another. During trials 100 through 300, the default XGBoost model achieves the lowest average runtime, the symbolic-profile model is second, and random search is highest. Shaded bands show 95 percent confidence intervals across thirty seeded runs.}
\end{figure}

Fig.~\ref{fig:convergence} shows the auto-tuner finds better kernels over trials. 
Fig.~\ref{fig:convergence-1to100} shows a large drop in kernel runtime over the first 100 trials, though the three methods on first 100 trials are very similar. 
It turns out the XGBoost framework uses a random model for the first 100 trials before the internal model is built and used. 

Fig.~\ref{fig:convergence-100to300} shows the kernel improvement over 100-300 trials. 
We can see that our model (the green line) is clearly better than the random model confirming that kernel profiles can predict kernel performance. 
For some seeds, the best achieved runtime of our model is comparable to xgb-default results. 
Overall, however, xgb-default provides the best results among the three. 
In addition, we can see that the shaded area of our model is larger than the other two, indicating larger variance and hinting that the instruction map features are  too coarse for prediction. 
Adding structural features about the location of instructions and memory-related features could result in better performance.

\begin{figure} %
    \centering
    \includegraphics[width=0.5\linewidth]{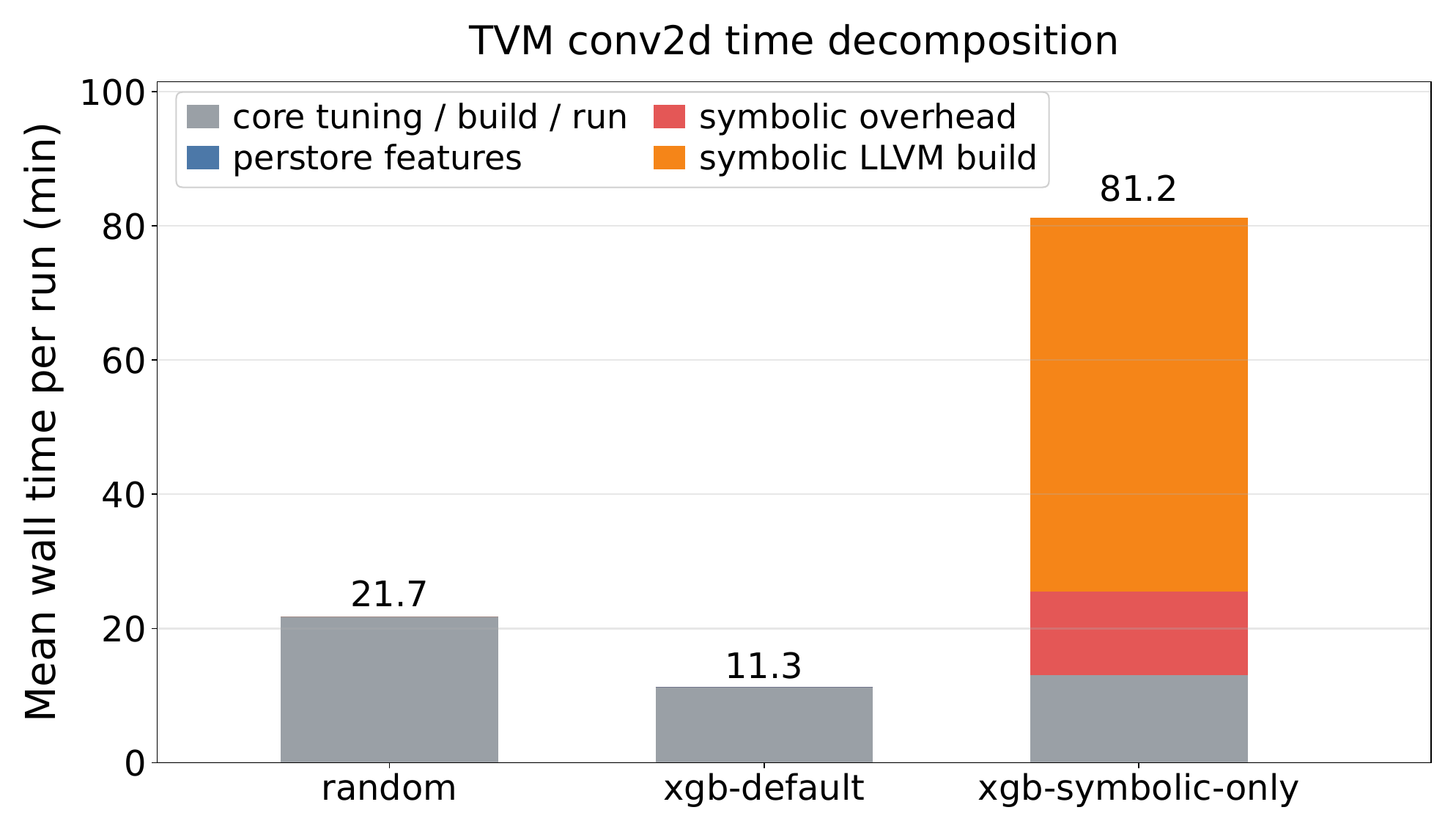}
    \caption{Total time decomposition. Each stack represents the average time to complete a 300-trials auto-tuning experiment using the method denoted below. The gray area is the time spent on model initialization, kernel generation and evaluation. The other three colored areas are time spent on feature extraction (perstore features are required by xgb-default, but the time is too small to be noticeable). }
    \label{fig:time-decomposition} 
    \Description{A stacked bar chart comparing the wall-clock time of three 300-trial auto-tuning methods. Random search takes 21.7 minutes and the default XGBoost method takes 11.3 minutes, primarily for core tuning, building, and execution. The symbolic-profile XGBoost method takes 81.2 minutes. Its much taller bar consists of core tuning time, symbolic-analysis overhead, and a dominant LLVM-building component; its feature-extraction component is negligible.}
\end{figure}

Fig.~\ref{fig:time-decomposition} shows the total time decomposition of the 300-trials experiment. The gray area represents basic auto-tuning overhead, which includes kernel generation, building and empirical evaluation.
The random method has larger gray area caused by the longer average evaluation time of proposed kernels (random model picks random kernels that can be slow). 
In comparison, both xgb-default and xgb-symbolic-only have similar area on the basic overhead, which indicates both methods evaluate better candidates. 

Fig.~\ref{fig:time-decomposition} also shows that our xgb-symbolic-only method has high additional overhead. 
While the red area represents the overhead of our symbolic analysis to extract formulae, most of the time is spent on compiling TVM IR down to LLVM IR (the orange area) since the `evolutionary' search proposes 64 candidates at each trial, and xgb-symbolic-only has to compile all the 64 kernels down to LLVM IR for analysis. 
In contrast, xgb-default derives its features directly from the TVM IR.

To view these overheads in the right context, it is important to understand that our symbolic analysis was designed to output reusable formulae to produce profiles for \textit{all} possible input sizes.
However, in TVM autotuning, input sizes are fixed for a kernel being autotuned (here 512x512) preventing us from amortizing the LLVM IR compilation across multiple input sizes.

\subsubsection{Limitations and Future Direction}
Although our experiments show that profiles can facilitate auto-tuning, their use is hamstrung in the current auto-tuning framework because of repeated LLVM IR lowering. 
A better way to apply symbolic profiling in auto-tuning would be to enable light-weight analysis across kernel optimizations. 
For example, kernels with only different tile sizes should get similar sets of symbolic formulae with tile sizes as a parameter, greatly reducing the number of required analyses since we no longer will treat each tiled kernel as a completely different kernel and run a full analysis on it. 
This would require reworking how autotuning works in most frameworks and therefore we leave incorporating symbolic profiling into auto-tuning effectively to future work.

\section{Related Work}\label{sec:related}

Program profiling aims to find the basic block counts of a program. Multiple methods can be used to find the counts, exactly or approximately, with different trade-offs. 
Traditional methods include instrumentation-based methods, approximate profiling, and sampling-based methods.
In each subsection, we compare our method to each category of such methods.

\subsection*{Exact Profiling (Instrumentation)}
Instrumentation inserts counters into a program to collect basic block counts during execution producing exact counts, but requires running the program.
This adds execution overhead, and thus a large corpus of research is focused on reducing the amount and performance impact of counters inserted.
\citet{knuthOptimalMeasurementPoints1973} described an algorithm to determine optimal counter placement. Based on this algorithm, \citet{OptimallyProfilingTracing} provided a general formulation for program profiling adopted by most current instrumentation-based profiling tools. \citet{ballEfficientPathProfiling1996} extended this approach to path profiling, which  differs from block profiling by further separating block counts for different paths.
Work to further optimize instrumentation-based techniques include \citet{frenotReducingOverheadExact2024} which improved general instrumentation-based profiling by reusing affine variables, and \citet{optimizingCustomizedProgramCoverage} proposed heuristics to reduce counters when profiling for coverage instead of exact counts. \citet{automaticPropagationOfProfileInformation} proposed methods to project existing profile information in PGO, avoiding the need to re-execute the optimized program to regenerate profiles. While these methods reduce the overhead of instrumentation, they fundamentally require running the program to collect basic block counts.

\subsection*{Approximate Profiling}
When exact counts are not necessary, approximate profiling can be used in place of instrumentation for significantly lower overhead. Sampling can yield approximate counts by periodically recording execution events to infer hot regions. 
\citet{aProgrammableHardwarePathProfiler} propose hardware-based profiling infrastructure capable of identifying \emph{hot paths} (i.e. frequently executed basic blocks) in a program, using a unique ``path descriptor'' that is more efficient to represent in hardware than basic blocks.
While it achieves low overhead hot path profiling, on realistic hardware it is unable to acquire exact basic block count due to resource constraints.

Empirical analytical models can discover approximate relationships between inputs and basic block counts.
\citet{goldsmithMeasuringEmpiricalComputational2007} provide a method to create empirical models of basic block counts in relation to their inputs. This method assumes basic blocks are in linear or power law relationship to input features, and fits a statistical model to sampled profiling data making it approximate and limited to fixed analytical models.

Static techniques \cite{calderEvidencebasedStaticBranch1997, moreiraVESPAStaticProfiling2021} can analyze the code to predict execution frequency without running the program.
However, these are based on machine learning over a corpus of programs and they do not produce exact profiles. For example, the model in \citet{moreiraVESPAStaticProfiling2021} only achieves an average of 80\% accuracy in predicting branch direction for a set of programs.

Some works focus on reconstructing the exact profile from the approximate counts \cite{profileInferenceRevisited, staticBranchFrequencyAndProgramProfileAnalysis}, enabling hybrid approaches that combine low-overhead profiling with high accuracy though these still do not provide exact counts. 

In some cases, empirical approximation can be very accurate and produce exact counts. 
Merlin \cite{sumitani-class-2023} produces empirical complexity measures and introduces "Single-Complexity Path" (SCP) programs whose complexity is a function of a single input feature.
While Merlin's SCP does admit naive ML kernels like \texttt{matmul}, it excludes \texttt{dropout} and \texttt{resize} where loops are controlled by boolean flags. In addition, optimizations break SCP, for example, vectorization creates residual loops whose execution depends on input shape. Our SDCK definition covers these well.

Despite those limitations, Merlin yields an accurate polynomial complexity model by identifying loops' depth to decide the order of target polynomials, and then using sampled profiling data to solve for parameters in the polynomial. 
The resultant model can be exact but is limited to single input variable. From a practical perspective, Merlin has been used to infer the asymptotic complexity of neural networks \cite{merlin-inferring-nn}, which is a possible use case of our technique as well.

\subsection*{Symbolic Program Analysis}
Symbolic methods reason about program behavior without execution, including extraction of performance-relevant properties. For example, symbolic analysis has been applied to extract closed-form expression of program execution cost, including time and other resources \cite{albert-closed-form-2011, mechanical-program-analysis}. \citet{speed} extract symbolic expressions that characterize the computational complexity of programs. \citet{stackSpaceBound} use symbolic reasoning to derive upper bounds on stack space usage, offering insights into memory consumption under different inputs. However, these works do not provide exact symbolic basic block profiles. 

\citet{rileyExactLoopBound2025} formulate the \emph{Exact Loop Bound Analysis} (ELBA) problem as a precondition synthesis problem, and use symbolic analysis to compute precise loop iteration counts as a function of program inputs.
This method generates symbolic models of loop conditions and branches using SMT solvers, and induces relations between program variables by unrolling models with specific initial conditions and fitting linear relationships.
Both this method and our method are able to handle multi-phase loops.
ELBA is limited to single-loop programs without nested loops, while ours works with multiple loops and nested loops.
ELBA supports more complex loop conditions due to the usage of invariant generation and verification, but this is computationally costly, incomplete (due to rejection from timeout), and unnecessary for ML kernels.

\subsection*{Profile-Guided Optimization}

Profile data has been used in impactful ways for compiler optimization~\cite{chenAutoFDOAutomaticFeedbackdirected2016, moreiraVESPAStaticProfiling2021, panchenkoBOLTPracticalBinary2019}. 
For instance, profiling data guides inlining decisions in data-center environments to reduce call overhead and improve locality \cite{chenAutoFDOAutomaticFeedbackdirected2016}. It also enables more accurate branch prediction strategies by exposing frequently taken paths \cite{songThermometerProfileGuidedBtbReplacement}, and helps improve instruction cache performance by informing code layout optimizations \cite{ panchenkoBOLTPracticalBinary2019}. 

\citet{chenAutoFDOAutomaticFeedbackdirected2016} and \citet{panchenkoBOLTPracticalBinary2019} sample basic block frequencies on large scale production applications by sampling last few taken branches and mapping them to instructions and intermediate representation of the binary, with hardware performance counter data taken from processes running in production. 
This method allows approximate basic block counting for complex software, but unlike our approach, it requires running processes, and does not yield symbolic representations of basic block counts.

\citet{songThermometerProfileGuidedBtbReplacement} (Thermometer) uses Intel PT to sample basic block execution traces, and uses such information to derive more advanced metrics for analytical purposes.
\citet{litzCrispCriticalSlicePrefetching} use DynamoRIO Memtrace or Intel PT to obtain execution traces.
Both methods require running a separate profiling pass, and neither provides exact basic block counts.

\section{Conclusion}\label{sec:conclusion}

We propose a novel symbolic method for program profiling that extracts symbolic expressions for execution counts of basic blocks of a program. Using a graph representation, we delimit the scope of application of our technique and describe using operational semantics the general approach to generate symbolic formulae for all instances derived from that representation. We implement the symbolic profiling technique as an LLVM analysis pass and show that our symbolic approach successfully captures the relation between inputs and basic blocks even for complex kernels and produces exact results. We also compare the speed of our technique to traditional instrumentation-based profiling, showing that our method can obtain basic block profile with a constant overhead across different sized inputs. 
This can provide significant speedup (up to 23,480,043,367$\times$ for large computation kernels) compared to dynamic instrumentation. 
We show how these now statically available profiles can be used to build performance models for auto-tuning.

\newpage
\section*{Data-Availability Statement}
For this work, an artifact~\cite{qiu2026symbolicartifact} exists. It contains three parts. The first is a tool that implements our symbolic profiling method, which accepts LLVM IR programs and outputs symbolic basic block counts. The second contains the set of 78 tested TVM kernels in LLVM IR and related benchmarking scripts. The third contains an integration of our tool into the TVM auto-tuning framework. The artifact allows generating all the figures and results in the paper to validate our claims.

\begin{acks}
We thank the anonymous reviewers for their careful reviews and valuable suggestions which significantly improved this work. We thank members of the systems group at the University of Rochester's Computer Science department for their comments, suggestions, and insightful discussions during the development of this work. This material is based upon work supported by the National Science Foundation under Award Nos.~2144384 and 2402806. Any opinions, findings and conclusions or recommendations expressed in this material are those of the author(s) and do not necessarily reflect the views of the National Science Foundation.
\end{acks}

\bibliographystyle{ACM-Reference-Format}
\bibliography{zot, manual, zot-ref}

@article{automaticPropagationOfProfileInformation,
author = {Fr{\"o}hlich, Elisa and Moreira, Ang{\'e}lica and Magno Quint{\~a}o Pereira, Fernando},
title = {Automatic Propagation of Profile Information through the Optimization Pipeline},
year = {2026},
issue_date = {April 2026},
publisher = {Association for Computing Machinery},
address = {New York, NY, USA},
volume = {10},
number = {OOPSLA1},
url = {https://doi.org/10.1145/3798247},
doi = {10.1145/3798247},
journal = {Proc. ACM Program. Lang.},
month = apr,
articleno = {139},
numpages = {26}
}

@misc{qiu2026symbolicartifact,
  author    = {Qiu, Jingyu and Dong, Rongcui and Pai, Sreepathi},
  title     = {Artifact for Symbolic Basic Block Profiling for Machine Learning Kernels},
  year      = {2026},
  publisher = {Zenodo},
  doi       = {10.5281/zenodo.21513685},
  url       = {https://doi.org/10.5281/zenodo.21513685}
}

@misc{merlin-inferring-nn,
  author       = {Rafael Fontes Sumitani and Lucas Victor da Silva Costa and Frederico F. Campos and Fernando Magno Quintão Pereira},
  title        = {Inferring the Asymptotic Complexity of Neural Networks
                  with {Merlin}},
  url = {https://github.com/lac-dcc/merlin/wiki/Inferring-the-Asymptotic-Complexity-of-Neural-Networks-with-Merlin},
  year         = {2023},
  lastaccessed = {July 16, 2026},
  note         = {GitHub Wiki}
}

@article{balaprakashAutotuningHighPerformanceComputing2018,
	title = {Autotuning in {High}-{Performance} {Computing} {Applications}},
	volume = {106},
	issn = {1558-2256},
	url = {https://ieeexplore.ieee.org/document/8423171/},
	doi = {10.1109/JPROC.2018.2841200},
	number = {11},
	urldate = {2025-09-19},
	journal = {Proceedings of the IEEE},
	author = {Balaprakash, Prasanna and Dongarra, Jack and Gamblin, Todd and Hall, Mary and Hollingsworth, Jeffrey K. and Norris, Boyana and Vuduc, Richard},
	month = nov,
	year = {2018},
	pages = {2068--2083},
}

@inproceedings{sumitani-class-2023,
	title = {A {Class} of {Programs} that {Admit} {Exact} {Complexity} {Analysis} via {Newton}'s {Polynomial} {Interpolation}},
	isbn = {979-8-4007-1628-7},
	shorttitle = {A {Class} of {Programs} that {Admit} {Exact} {Complexity} {Analysis} via {Newton}?},
	url = {https://dl.acm.org/doi/10.1145/3624309.3624311},
	doi = {10.1145/3624309.3624311},
	language = {en},
	urldate = {2026-05-20},
	booktitle = {Proceedings of the {XXVII} {Brazilian} {Symposium} on {Programming} {Languages}},
	publisher = {Association for Computing Machinery},
    address   = {New York, NY, USA},
    location  = {Campo Grande, MS, Brazil},
	author = {Rafael Fontes Sumitani and
          Lucas Victor da Silva Costa and
          Frederico F. Campos and
          Fernando Magno Quintão Pereira},
	month = sep,
	year = {2023},
	pages = {50--55},
}

@misc{onnxOperators,
  title        = {ONNX Operators},
  author       = {{ONNX Community}},
  url = {https://onnx.ai/onnx/operators/},
  year         = 2026,
  lastaccessed = {March 17, 2026}
}

@INPROCEEDINGS{polygeist,
  author={Moses, William S. and Chelini, Lorenzo and Zhao, Ruizhe and Zinenko, Oleksandr},
  booktitle={2021 30th International Conference on Parallel Architectures and Compilation Techniques (PACT)}, 
  location = {Atlanta, GA, USA},
  title={Polygeist: Raising C to Polyhedral MLIR}, 
  year={2021},
  volume={},
  number={},
  pages={45-59},
  doi={10.1109/PACT52795.2021.00011}}

@misc{tvmTutorialE2E,
  author       = {{Apache TVM Community}},
  title        = {End-to-End Optimize Model Tutorial},
  howpublished = {Online tutorial, Apache TVM documentation},
  year         = {2024},
  lastaccessed = {November 13, 2025},
  url          = {https://tvm.apache.org/docs/how_to/tutorials/e2e_opt_model.html}
}

@inproceedings{tensorFlow,
author = {Abadi, Mart\'{\i}n and Barham, Paul and Chen, Jianmin and Chen, Zhifeng and Davis, Andy and Dean, Jeffrey and Devin, Matthieu and Ghemawat, Sanjay and Irving, Geoffrey and Isard, Michael and Kudlur, Manjunath and Levenberg, Josh and Monga, Rajat and Moore, Sherry and Murray, Derek G. and Steiner, Benoit and Tucker, Paul and Vasudevan, Vijay and Warden, Pete and Wicke, Martin and Yu, Yuan and Zheng, Xiaoqiang},
title = {TensorFlow: a system for large-scale machine learning},
year = {2016},
isbn = {9781931971331},
publisher = {USENIX Association},
address = {USA},
booktitle = {Proceedings of the 12th USENIX Conference on Operating Systems Design and Implementation},
pages = {265–283},
numpages = {19},
location = {Savannah, GA, USA},
series = {OSDI'16}
}

@article{mechanical-program-analysis,
author = {Wegbreit, Ben},
title = {Mechanical program analysis},
year = {1975},
issue_date = {Sept. 1975},
publisher = {Association for Computing Machinery},
address = {New York, NY, USA},
volume = {18},
number = {9},
issn = {0001-0782},
url = {https://doi.org/10.1145/361002.361016},
doi = {10.1145/361002.361016},
journal = {Commun. ACM},
month = sep,
pages = {528–539},
numpages = {12}
}

@article{albert-closed-form-2011,
	title = {Closed-{Form} {Upper} {Bounds} in {Static} {Cost} {Analysis}},
	volume = {46},
	copyright = {http://www.springer.com/tdm},
	issn = {0168-7433, 1573-0670},
	url = {http://link.springer.com/10.1007/s10817-010-9174-1},
	doi = {10.1007/s10817-010-9174-1},
	language = {en},
	number = {2},
	urldate = {2025-06-24},
	journal = {Journal of Automated Reasoning},
	author = {Albert, Elvira and Arenas, Puri and Genaim, Samir and Puebla, Germán},
	month = feb,
	year = {2011},
	pages = {161--203},
}

@misc{onnx-model-zoo,
  author       = {{The ONNX Contributors}},
  title        = {ONNX Model Zoo},
  url = {https://onnx.ai/models/},
  year         = {2025},
  lastaccessed = {September 10, 2025}
}

@inproceedings {TVM,
author = {Tianqi Chen and Thierry Moreau and Ziheng Jiang and Lianmin Zheng and Eddie Yan and Haichen Shen and Meghan Cowan and Leyuan Wang and Yuwei Hu and Luis Ceze and Carlos Guestrin and Arvind Krishnamurthy},
title = {{TVM}: An Automated {End-to-End} Optimizing Compiler for Deep Learning},
booktitle = {13th USENIX Symposium on Operating Systems Design and Implementation (OSDI 18)},
year = {2018},
isbn = {978-1-939133-08-3},
location = {Carlsbad, CA, USA},
pages = {578--594},
url = {https://www.usenix.org/conference/osdi18/presentation/chen},
publisher = {USENIX Association},
lastaccessed = {June 17, 2026},
month = oct
}

@inproceedings{speed,
author = {Gulwani, Sumit and Mehra, Krishna K. and Chilimbi, Trishul},
title = {SPEED: precise and efficient static estimation of program computational complexity},
year = {2009},
isbn = {9781605583792},
publisher = {Association for Computing Machinery},
address = {New York, NY, USA},
url = {https://doi.org/10.1145/1480881.1480898},
doi = {10.1145/1480881.1480898},
booktitle = {Proceedings of the 36th Annual ACM SIGPLAN-SIGACT Symposium on Principles of Programming Languages},
pages = {127–139},
numpages = {13},
location = {Savannah, GA, USA},
series = {POPL '09}
}

@inproceedings{demouraZ3EfficientSMT2008,
  title = {Z3: {{An Efficient SMT Solver}}},
  shorttitle = {Z3},
  booktitle = {Tools and {{Algorithms}} for the {{Construction}} and {{Analysis}} of {{Systems}}},
  author = {{de Moura}, Leonardo and Bj{\o}rner, Nikolaj},
  editor = {Ramakrishnan, C. R. and Rehof, Jakob},
  year = {2008},
  pages = {337--340},
  publisher = {Springer},
  address = {Berlin, Heidelberg},
  doi = {10.1007/978-3-540-78800-3_24},
  isbn = {978-3-540-78800-3},
  langid = {english}
}

@article{rileyExactLoopBound2025,
  author = {Riley, Daniel and Fedyukovich, Grigory},
title = {Exact Loop Bound Analysis},
year = {2025},
issue_date = {June 2025},
publisher = {Association for Computing Machinery},
address = {New York, NY, USA},
volume = {9},
number = {PLDI},
url = {https://doi.org/10.1145/3729323},
doi = {10.1145/3729323},
journal = {Proc. ACM Program. Lang.},
month = jun,
articleno = {220},
numpages = {24}
}

@inproceedings{goldsmithMeasuringEmpiricalComputational2007,
  author = {Goldsmith, Simon F. and Aiken, Alex S. and Wilkerson, Daniel S.},
title = {Measuring empirical computational complexity},
year = {2007},
isbn = {9781595938114},
publisher = {Association for Computing Machinery},
address = {New York, NY, USA},
url = {https://doi.org/10.1145/1287624.1287681},
doi = {10.1145/1287624.1287681},
booktitle = {Proceedings of the 6th Joint Meeting of the European Software Engineering Conference and the ACM SIGSOFT Symposium on The Foundations of Software Engineering},
pages = {395–404},
numpages = {10},
location = {Dubrovnik, Croatia},
series = {ESEC-FSE '07}
}

@misc{plasmaUmassBigO,
    author       = {Berger, Emery},
  title        = {{bigO}: Measures Empirical Computational Complexity in Time and Space},
  year         = {2026},
  url = {https://github.com/plasma-umass/bigO},
  lastaccessed = {June 30, 2026},
  note         = {Version v0.0.8}
}

@inproceedings{anselOpenTunerExtensibleFramework2014,
    title = {{OpenTuner}: {An} extensible framework for program autotuning},
    shorttitle = {{OpenTuner}},
    url = {https://ieeexplore.ieee.org/document/7855909/},
    doi = {10.1145/2628071.2628092},
    urldate = {2025-09-19},
    booktitle = {2014 23rd {International} {Conference} on {Parallel} {Architecture} and {Compilation} {Techniques} ({PACT})},
    author = {Ansel, Jason and Kamil, Shoaib and Veeramachaneni, Kalyan and Ragan-Kelley, Jonathan and Bosboom, Jeffrey and O'Reilly, Una-May and Amarasinghe, Saman},
    month = aug,
    year = {2014},
    location = {Edmonton, AB, Canada},
    pages = {303--316},
}

@inproceedings{schulzePyATFConstraintBasedAutoTuning2025,
    location = {Las Vegas, NV, USA},
    series = {CC '25},
    title = {{pyATF}: {Constraint}-{Based} {Auto}-{Tuning} in {Python}},
    isbn = {979-8-4007-1407-8},
    shorttitle = {{pyATF}},
    url = {https://dl.acm.org/doi/10.1145/3708493.3712682},
    doi = {10.1145/3708493.3712682},
    urldate = {2026-01-14},
    booktitle = {Proceedings of the 34th {ACM} {SIGPLAN} {International} {Conference} on {Compiler} {Construction}},
    publisher = {Association for Computing Machinery},
    address   = {New York, NY, USA},
    author = {Schulze, Richard and Gorlatch, Sergei and Rasch, Ari},
    month = feb,
    year = {2025},
    pages = {35--47},
}

@inproceedings{chen-xgboost-2016,
    title = {{XGBoost}: {A} {Scalable} {Tree} {Boosting} {System}},
    shorttitle = {{XGBoost}},
    url = {http://arxiv.org/abs/1603.02754},
    doi = {10.1145/2939672.2939785},
    urldate = {2026-07-05},
    booktitle = {Proceedings of the 22nd {ACM} {SIGKDD} {International} {Conference} on {Knowledge} {Discovery} and {Data} {Mining}},
    author = {Chen, Tianqi and Guestrin, Carlos},
    month = aug,
    year = {2016},
    pages = {785--794},
    location = {San Francisco, CA, USA},
}

@inproceedings{litzCrispCriticalSlicePrefetching,
author = {Litz, Heiner and Ayers, Grant and Ranganathan, Parthasarathy},
title = {CRISP: critical slice prefetching},
year = {2022},
isbn = {9781450392051},
publisher = {Association for Computing Machinery},
address = {New York, NY, USA},
url = {https://doi.org/10.1145/3503222.3507745},
doi = {10.1145/3503222.3507745},
booktitle = {Proceedings of the 27th ACM International Conference on Architectural Support for Programming Languages and Operating Systems},
pages = {300–313},
numpages = {14},
location = {Lausanne, Switzerland},
series = {ASPLOS '22}
}

@inproceedings{songThermometerProfileGuidedBtbReplacement,
author = {Song, Shixin and Khan, Tanvir Ahmed and Shahri, Sara Mahdizadeh and Sriraman, Akshitha and Soundararajan, Niranjan K and Subramoney, Sreenivas and Jim\'{e}nez, Daniel A. and Litz, Heiner and Kasikci, Baris},
title = {Thermometer: profile-guided {BTB} replacement for data center applications},
year = {2022},
isbn = {9781450386104},
publisher = {Association for Computing Machinery},
address = {New York, NY, USA},
url = {https://doi.org/10.1145/3470496.3527430},
doi = {10.1145/3470496.3527430},
booktitle = {Proceedings of the 49th Annual International Symposium on Computer Architecture},
pages = {742–756},
numpages = {15},
location = {New York, New York},
series = {ISCA '22}
}

@INPROCEEDINGS{soaresMemorySafeElimination,
  author = {Soares, Luigi and Pereira, Fernando Magno Quint{\~a}o},
  booktitle={2021 IEEE/ACM International Symposium on Code Generation and Optimization (CGO)}, 
  title={Memory-Safe Elimination of Side Channels}, 
  year={2021},
  volume={},
  number={},
  location = {Virtual Event, Republic of Korea},
  pages={200-210},
  doi={10.1109/CGO51591.2021.9370305}}

@article{soaresSideChannelElimination,
author = {Soares, Luigi and Canesche, Michael and Pereira, Fernando Magno Quint\~{a}o},
title = {Side-channel Elimination via Partial Control-flow Linearization},
year = {2023},
issue_date = {June 2023},
publisher = {Association for Computing Machinery},
address = {New York, NY, USA},
volume = {45},
number = {2},
issn = {0164-0925},
url = {https://doi.org/10.1145/3594736},
doi = {10.1145/3594736},
journal = {ACM Trans. Program. Lang. Syst.},
month = jun,
articleno = {13},
numpages = {43}
}

@ARTICLE{chenTamingHardwareEventSamples,
  author={Chen, Dehao and Vachharajani, Neil and Hundt, Robert and Li, Xinliang and Eranian, Stephane and Chen, Wenguang and Zheng, Weimin},
  journal={IEEE Transactions on Computers}, 
  title={Taming Hardware Event Samples for Precise and Versatile Feedback Directed Optimizations}, 
  year={2013},
  volume={62},
  number={2},
  pages={376-389},
  doi={10.1109/TC.2011.233}}

@inproceedings{stackSpaceBound,
author = {Carbonneaux, Quentin and Hoffmann, Jan and Ramananandro, Tahina and Shao, Zhong},
title = {End-to-end verification of stack-space bounds for C programs},
year = {2014},
isbn = {9781450327848},
publisher = {Association for Computing Machinery},
address = {New York, NY, USA},
url = {https://doi.org/10.1145/2594291.2594301},
doi = {10.1145/2594291.2594301},
booktitle = {Proceedings of the 35th ACM SIGPLAN Conference on Programming Language Design and Implementation},
pages = {270–281},
numpages = {12},
location = {Edinburgh, United Kingdom},
series = {PLDI '14}
}

@article{profileInferenceRevisited,
author = {He, Wenlei and Mestre, Juli\'{a}n and Pupyrev, Sergey and Wang, Lei and Yu, Hongtao},
title = {Profile inference revisited},
year = {2022},
issue_date = {January 2022},
publisher = {Association for Computing Machinery},
address = {New York, NY, USA},
volume = {6},
number = {POPL},
url = {https://doi.org/10.1145/3498714},
doi = {10.1145/3498714},
journal = {Proc. ACM Program. Lang.},
month = jan,
articleno = {52},
numpages = {24}
}

@inproceedings{staticBranchFrequencyAndProgramProfileAnalysis,
author = {Wu, Youfeng and Larus, James R.},
title = {Static branch frequency and program profile analysis},
year = {1994},
isbn = {0897917073},
publisher = {Association for Computing Machinery},
address = {New York, NY, USA},
url = {https://doi.org/10.1145/192724.192725},
doi = {10.1145/192724.192725},
booktitle = {Proceedings of the 27th Annual International Symposium on Microarchitecture},
pages = {1–11},
numpages = {11},
location = {San Jose, California, USA},
series = {MICRO 27}
}

@inproceedings{optimizingCustomizedProgramCoverage,
author = {Ohmann, Peter and Brown, David Bingham and Neelakandan, Naveen and Linderoth, Jeff and Liblit, Ben},
title = {Optimizing customized program coverage},
year = {2016},
isbn = {9781450338455},
publisher = {Association for Computing Machinery},
address = {New York, NY, USA},
url = {https://doi.org/10.1145/2970276.2970351},
doi = {10.1145/2970276.2970351},
booktitle = {Proceedings of the 31st IEEE/ACM International Conference on Automated Software Engineering},
pages = {27–38},
numpages = {12},
location = {Singapore, Singapore},
series = {ASE '16}
}

@INPROCEEDINGS{aProgrammableHardwarePathProfiler,
  author={Kapil Vaswani and Thazhuthaveetil, M.J. and Srikant, Y.N.},
  booktitle={International Symposium on Code Generation and Optimization}, 
  title={A programmable hardware path profiler}, 
  location = {San Jose, CA, USA},
  year={2005},
  volume={},
  number={},
  pages={217-228},
  doi={10.1109/CGO.2005.3}}

@INPROCEEDINGS{continuousPathAndEdgeProfiling,
  author = {Bond, Michael D. and McKinley, Kathryn S.},
    title = {Continuous Path and Edge Profiling},
    year = {2005},
    isbn = {0769524400},
    publisher = {IEEE Computer Society},
    address = {USA},
    url = {https://doi.org/10.1109/MICRO.2005.16},
    doi = {10.1109/MICRO.2005.16},
    booktitle = {Proceedings of the 38th Annual IEEE/ACM International Symposium on Microarchitecture},
    pages = {130–140},
    numpages = {11},
    location = {Barcelona, Spain},
    series = {MICRO 38}
}

@article{usingProfileInformation,
author = {Chang, Pohua P. and Mahlke, Scott A. and Hwu, Wen-Mei W.},
title = {Using profile information to assist classic code optimizations},
journal = {Software: Practice and Experience},
volume = {21},
number = {12},
pages = {1301-1321},
doi = {https://doi.org/10.1002/spe.4380211204},
url = {https://onlinelibrary.wiley.com/doi/abs/10.1002/spe.4380211204},
year = {1991}
}

@inproceedings{ballEfficientPathProfiling1996,
  title = {Efficient Path Profiling},
  booktitle = {Proceedings of the 29th Annual IEEE/ACM International
             Symposium on Microarchitecture (MICRO 29)},
  author = {Thomas Ball and James R. Larus},
  year = {1996},
  pages = {46--57},
  publisher = {IEEE Comput. Soc. Press},
  location = {Paris, France},
  doi = {10.1109/MICRO.1996.566449},
  urldate = {2024-12-30},
  isbn = {978-0-8186-7641-3},
  langid = {english}
}

@article{calderEvidencebasedStaticBranch1997,
  title = {Evidence-Based Static Branch Prediction Using Machine Learning},
  author = {Calder, Brad and Grunwald, Dirk and Jones, Michael and Lindsay, Donald and Martin, James and Mozer, Michael and Zorn, Benjamin},
  year = {1997},
  month = jan,
  journal = {ACM Transactions on Programming Languages and Systems},
  volume = {19},
  number = {1},
  pages = {188--222},
  issn = {0164-0925, 1558-4593},
  doi = {10.1145/239912.239923},
  urldate = {2025-01-06},
  langid = {english}
}

@inproceedings{chenAutoFDOAutomaticFeedbackdirected2016,
  title = {{{AutoFDO}}: Automatic Feedback-Directed Optimization for Warehouse-Scale Applications},
  shorttitle = {{{AutoFDO}}},
  booktitle = {Proceedings of the 2016 {{International Symposium}} on {{Code Generation}} and {{Optimization}}},
  author = {Chen, Dehao and Li, David Xinliang and Moseley, Tipp},
  year = {2016},
  month = feb,
  pages = {12--23},
  publisher = {Association for Computing Machinery},
  address   = {New York, NY, USA},
  location  = {Barcelona, Spain},
  doi = {10.1145/2854038.2854044},
  urldate = {2025-01-06},
  isbn = {978-1-4503-3778-6},
  langid = {english}
}

@inproceedings{frenotReducingOverheadExact2024,
  title = {Reducing the {{Overhead}} of {{Exact Profiling}} by {{Reusing Affine Variables}}},
  booktitle = {Proceedings of the 33rd {{ACM SIGPLAN International Conference}} on {{Compiler Construction}}},
  author = {Frenot, Leon and Pereira, Fernando Magno Quint{\~a}o},
  year = {2024},
  month = feb,
  pages = {150--161},
  publisher = {Association for Computing Machinery},
  address   = {New York, NY, USA},
  location  = {Edinburgh, United Kingdom},
  doi = {10.1145/3640537.3641569},
  urldate = {2024-12-31},
  isbn = {979-8-4007-0507-6},
  langid = {english}
}

@article{knuthOptimalMeasurementPoints1973,
  title = {Optimal Measurement Points for Program Frequency Counts},
  author = {Knuth, Donald E. and Stevenson, Francis R.},
  year = {1973},
  month = sep,
  journal = {BIT},
  volume = {13},
  number = {3},
  pages = {313--322},
  issn = {0006-3835, 1572-9125},
  doi = {10.1007/BF01951942},
  urldate = {2024-12-31},
  copyright = {http://www.springer.com/tdm},
  langid = {english}
}

@article{moreiraVESPAStaticProfiling2021,
  author = {Moreira, Ang{\'e}lica Aparecida and Ottoni, Guilherme and Quint{\~a}o Pereira, Fernando Magno},
title = {VESPA: static profiling for binary optimization},
year = {2021},
issue_date = {October 2021},
publisher = {Association for Computing Machinery},
address = {New York, NY, USA},
volume = {5},
number = {OOPSLA},
url = {https://doi.org/10.1145/3485521},
doi = {10.1145/3485521},
journal = {Proc. ACM Program. Lang.},
month = oct,
articleno = {144},
numpages = {28}
}

@article{OptimallyProfilingTracing,
author = {Ball, Thomas and Larus, James R.},
title = {Optimally profiling and tracing programs},
year = {1994},
issue_date = {July 1994},
publisher = {Association for Computing Machinery},
address = {New York, NY, USA},
volume = {16},
number = {4},
issn = {0164-0925},
url = {https://doi.org/10.1145/183432.183527},
doi = {10.1145/183432.183527},
journal = {ACM Trans. Program. Lang. Syst.},
month = jul,
pages = {1319–1360},
numpages = {42}
}

@inproceedings{panchenkoBOLTPracticalBinary2019,
  title = {{{BOLT}}: {{A Practical Binary Optimizer}} for {{Data Centers}} and {{Beyond}}},
  shorttitle = {{{BOLT}}},
  booktitle = {2019 {{IEEE}}/{{ACM International Symposium}} on {{Code Generation}} and {{Optimization}} ({{CGO}})},
  author = {Panchenko, Maksim and Auler, Rafael and Nell, Bill and Ottoni, Guilherme},
  year = {2019},
  month = feb,
  pages = {2--14},
  publisher = {IEEE},
  location = {Washington, DC, USA},
  doi = {10.1109/CGO.2019.8661201},
  urldate = {2025-01-02},
  copyright = {https://ieeexplore.ieee.org/Xplorehelp/downloads/license-information/IEEE.html},
  isbn = {978-1-7281-1436-1},
  langid = {english}
}

@inproceedings{paszkePyTorchImperativeStyle2019,
  author = {Paszke, Adam and Gross, Sam and Massa, Francisco and Lerer, Adam and Bradbury, James and Chanan, Gregory and Killeen, Trevor and Lin, Zeming and Gimelshein, Natalia and Antiga, Luca and Desmaison, Alban and Kopf, Andreas and Yang, Edward and DeVito, Zachary and Raison, Martin and Tejani, Alykhan and Chilamkurthy, Sasank and Steiner, Benoit and Fang, Lu and Bai, Junjie and Chintala, Soumith},
 title = {{PyTorch}: An Imperative Style, High-Performance Deep Learning Library},
  booktitle = {Advances in Neural Information Processing Systems},
  volume = {32},
  pages = {8026--8037},
  year = {2019},
  location = {Vancouver, BC, Canada},
 url = {https://proceedings.neurips.cc/paper_files/paper/2019/file/bdbca288fee7f92f2bfa9f7012727740-Paper.pdf},
}
\end{document}